\documentstyle[bezier]{article}
\makeatletter

\def\[{\relax\ifmmode\@badmath\else
 \begin{trivlist}%
 \@beginparpenalty\predisplaypenalty
 \@endparpenalty\postdisplaypenalty
 \item[]\leavevmode
 \hbox to\linewidth\bgroup $\m@th\displaystyle
 \hskip\mathindent\bgroup\fi}

\def\]{\relax\ifmmode \egroup $\hfil
       \egroup \end{trivlist}\else \@badmath \fi}

\def\equation{\@beginparpenalty\predisplaypenalty
  \@endparpenalty\postdisplaypenalty
\refstepcounter{equation}\trivlist \item[]\leavevmode
  \hbox to\linewidth\bgroup $\m@th
  \displaystyle
\hskip\mathindent}

\def\endequation{$\hfil
           \displaywidth\linewidth\@eqnnum\egroup \endtrivlist}

\def\eqnarray{\stepcounter{equation}\let\@currentlabel=\theequation
\global\@eqnswtrue
\global\@eqcnt\z@\tabskip\mathindent\let\\=\@eqncr
\abovedisplayskip\topsep\ifvmode\advance\abovedisplayskip\partopsep\fi
\belowdisplayskip\abovedisplayskip
\belowdisplayshortskip\abovedisplayskip
\abovedisplayshortskip\abovedisplayskip
$$\m@th\halign
to\linewidth\bgroup\@eqnsel\hskip\@centering$\displaystyle\tabskip\z@
  {##}$&\global\@eqcnt\@ne \hskip 2\arraycolsep \hfil${##}$\hfil
  &\global\@eqcnt\tw@ \hskip 2\arraycolsep $\displaystyle{##}$\hfil
   \tabskip\@centering&\llap{##}\tabskip\z@\cr}

\def\endeqnarray{\@@eqncr\egroup
      \global\advance\c@equation\m@ne$$\global\@ignoretrue
      }

\newdimen\mathindent
\mathindent = \leftmargini

\newdimen\@bls                    
\@bls=\baselineskip               
\advance\@bls -1ex                
\newdimen\@eps                    %
\def\section{\@startsection{section}{1}{\z@}
  {1.5\@bls plus 0.5\@bls}{1\@bls}{\normalsize\bf}}
\def\subsection{\@startsection{subsection}{2}{\z@}
  {1\@bls plus 0.25\@bls}{\@eps}{\normalsize\bf}}
\def\subsubsection{\@startsection{subsubsection}{3}{\z@}
  {1\@bls plus 0.25\@bls}{\@eps}{\normalsize\bf}}
\def\paragraph{\@startsection{paragraph}{4}{\parindent}
  {1\@bls plus 0.25\@bls}{0.5em}{\normalsize\bf}}
\def\subparagraph{\@startsection{subparagraph}{4}{\parindent}
  {1\@bls plus 0.25\@bls}{0.5em}{\normalsize\bf}}

\def\@sect#1#2#3#4#5#6[#7]#8{\ifnum #2>\c@secnumdepth
  \def\@svsec{}\else
  \refstepcounter{#1}\edef\@svsec{\csname the#1\endcsname.\hskip0.5em}\fi
  \@tempskipa #5\relax
  \ifdim \@tempskipa>\z@
    \begingroup
      #6\relax
      \@hangfrom{\hskip #3\relax\@svsec}{\interlinepenalty \@M #8\par}%
    \endgroup
    \csname #1mark\endcsname{#7}\addcontentsline
      {toc}{#1}{\ifnum #2>\c@secnumdepth \else
        \protect\numberline{\csname the#1\endcsname}\fi #7}%
  \else
    \def\@svsechd{#6\hskip #3\@svsec #8\csname #1mark\endcsname
      {#7}\addcontentsline{toc}{#1}{\ifnum #2>\c@secnumdepth \else
        \protect\numberline{\csname the#1\endcsname}\fi #7}}%
  \fi \@xsect{#5}}

\long\def\@makefigurecaption#1#2{\vskip 10mm #1. #2\par}

\long\def\@maketablecaption#1#2{\hbox to \hsize{\parbox[t]{\hsize}
  {#1 \\ #2}}\vskip 0.3ex}

\def\fnum@figure{Figure \thefigure}
\def\figure{\let\@makecaption\@makefigurecaption \@float{figure}}
\@namedef{figure*}{\let\@makecaption\@makefigurecaption \@dblfloat{figure}}

\def\table{\let\@makecaption\@maketablecaption \@float{table}}
\@namedef{table*}{\let\@makecaption\@maketablecaption \@dblfloat{table}}

\textfloatsep=\floatsep           
\intextsep=\floatsep              

\long\def\@makefntext#1{\parindent 1em\noindent\hbox{${}^{\@thefnmark}$}#1}

\def\maketitle{\begingroup        
    \def\thefootnote{\fnsymbol{footnote}}%
    \newpage \global\@topnum\z@
    \@maketitle \@thanks
  \endgroup
  \let\maketitle\relax \let\@maketitle\relax
  \gdef\@thanks{}\let\thanks\relax
  \gdef\@address{}\gdef\@author{}\gdef\@title{}\let\address\relax}

\def\justify@on{\let\\=\@normalcr
  \leftskip\z@ \@rightskip\z@ \rightskip\@rightskip}

\newbox\fm@box                    

\def\@maketitle{
  \global\setbox\fm@box=\vbox\bgroup
    \vskip0mm                    
    \raggedright                  
    \hyphenpenalty\@M             
    {\Large \@title \par}         
    \vskip\@bls                   
    {\normalsize                  
     \@author \par}               
    \vskip\@bls                   
    \@address                     
  \egroup
  \twocolumn[
    \unvbox\fm@box                
    \vskip\@bls                   
    \unvbox\abstract@box          
    \vskip 2pc]}                  

\newcounter{address}
\def\theaddress{\alph{address}}
\def\@makeadmark#1{\hbox{$^{\rm #1}$}}

\def\address#1{\addressmark\begingroup
  \xdef\@tempa{\theaddress}\let\\=\relax
  \def\protect{\noexpand\protect\noexpand}\xdef\@address{\@address
  \protect\addresstext{\@tempa}{#1}}\endgroup}
\def\@address{}

\def\addressmark{\stepcounter{address}%
  \xdef\@tempb{\theaddress}\@makeadmark{\@tempb}}

\def\addresstext#1#2{\leavevmode \begingroup
  \raggedright \hyphenpenalty\@M \@makeadmark{#1}#2\par \endgroup
  \vskip\@bls}

\newbox\abstract@box              

\def\abstract{%
  \global\setbox\abstract@box=\vbox\bgroup
  \small\rm
  \ignorespaces}
\def\endabstract{\par \egroup}

\def\thebibliography#1{\section*{REFERENCES}\list{\arabic{enumi}.}
  {\settowidth\labelwidth{#1.}\leftmargin=1.67em
   \labelsep\leftmargin \advance\labelsep-\labelwidth
   \itemsep\z@ \parsep\z@
   \usecounter{enumi}}\def\makelabel##1{\rlap{##1}\hss}%
   \def\newblock{\hskip 0.11em plus 0.33em minus -0.07em}
   \sloppy \clubpenalty=4000 \widowpenalty=4000 \sfcode`\.=1000\relax}

\newcount\@tempcntc
\def\@citex[#1]#2{\if@filesw\immediate\write\@auxout{\string\citation{#2}}\fi
  \@tempcnta\z@\@tempcntb\m@ne\def\@citea{}\@cite{\@for\@citeb:=#2\do
    {\@ifundefined
       {b@\@citeb}{\@citeo\@tempcntb\m@ne\@citea
        \def\@citea{,\penalty\@m\ }{\bf ?}\@warning
       {Citation `\@citeb' on page \thepage \space undefined}}%
    {\setbox\z@\hbox{\global\@tempcntc0\csname b@\@citeb\endcsname\relax}%
     \ifnum\@tempcntc=\z@ \@citeo\@tempcntb\m@ne
       \@citea\def\@citea{,\penalty\@m}
       \hbox{\csname b@\@citeb\endcsname}%
     \else
      \advance\@tempcntb\@ne
      \ifnum\@tempcntb=\@tempcntc
      \else\advance\@tempcntb\m@ne\@citeo
      \@tempcnta\@tempcntc\@tempcntb\@tempcntc\fi\fi}}\@citeo}{#1}}

\def\@citeo{\ifnum\@tempcnta>\@tempcntb\else\@citea
  \def\@citea{,\penalty\@m}%
  \ifnum\@tempcnta=\@tempcntb\the\@tempcnta\else
   {\advance\@tempcnta\@ne\ifnum\@tempcnta=\@tempcntb \else
\def\@citea{--}\fi
    \advance\@tempcnta\m@ne\the\@tempcnta\@citea\the\@tempcntb}\fi\fi}

\def\ps@crcplain{\let\@mkboth\@gobbletwo
     \def\@oddhead{\reset@font{\sl\rightmark}\hfil \rm\thepage}%
     \def\@evenhead{\reset@font\rm \thepage\hfil\sl\leftmark}%
     \let\@oddfoot\@empty
     \let\@evenfoot\@oddfoot}

\ps@crcplain                    
\makeatother

\def\FKS{F_{\rm KS}}

\def\setfonts{%
\font\frbig=eufm10
\font\frscr=eufm8 scaled\magstephalf
\font\frscrscr=eufm8
\newfam\frfam
\textfont\frfam=\frbig
\scriptfont\frfam=\frscr
\scriptscriptfont\frfam=\frscrscr
\def\fr{\fam\frfam}

\font\openbig=msbm10
\font\openscr=msbm8 scaled\magstephalf
\font\openscrscr=msbm8
\newfam\openfam
\textfont\openfam=\openbig
\scriptfont\openfam=\openscr
\scriptscriptfont\openfam=\openscrscr
\def\open{\fam\openfam}

\font\ssfbig=cmss10
\font\ssfscr=cmss8 
\font\ssfscrscr=cmss8
\newfam\ssffam
\textfont\ssffam=\ssfbig
\scriptfont\ssffam=\ssfscr
\scriptscriptfont\ssffam=\ssfscrscr
\def\ssf{\fam\ssffam}
}

\makeatletter
\@addtoreset{equation}{section}
\newdimen\normalarrayskip
\newdimen\minarrayskip
\normalarrayskip\baselineskip
\minarrayskip\jot
\newif\ifold \oldtrue \def\new{\oldfalse}
\def\arraymode{\ifold\relax\else\displaystyle\fi}

\def\@arrayskip{\ifold\baselineskip\z@\lineskip\z@
  \else
  \baselineskip\minarrayskip\lineskip2\minarrayskip\fi}
\def\@arrayclassz{\ifcase \@lastchclass \@acolampacol \or
\@ampacol \or \or \or \@addamp \or
 \@acolampacol \or \@firstampfalse \@acol \fi
\edef\@preamble{\@preamble
 \ifcase \@chnum
  \hfil$\relax\arraymode\@sharp$\hfil
  \or $\relax\arraymode\@sharp$\hfil
  \or \hfil$\relax\arraymode\@sharp$\fi}}
\def\@array[#1]#2{\setbox\@arstrutbox=\hbox{\vrule
  height\arraystretch \ht\strutbox
  depth\arraystretch \dp\strutbox
  width\z@}\@mkpream{#2}\edef\@preamble{\halign \noexpand\@halignto
\bgroup \tabskip\z@ \@arstrut \@preamble \tabskip\z@ \cr}%
\let\@startpbox\@@startpbox \let\@endpbox\@@endpbox
 \if #1t\vtop \else \if#1b\vbox \else \vcenter \fi\fi
 \bgroup \let\par\relax
 \let\@sharp##\let\protect\relax
 \@arrayskip\@preamble}
\@addtoreset{equation}{section}
\makeatother

\def\theequation{\thesection.\arabic{equation}}
\setfonts

\def\req#1{(\ref{#1})}

\def\BE{\begin{equation}}
\def\EE{\end{equation} }
\def\BA{\begin{array}} 
\def\EA{\end{array}}
\def\L{\left}
\def\R{\right}

\def\bar{\overline}
\def\frac#1#2{\mathchoice{{%
\textstyle{{#1}\over{#2}}}}{{#1\over#2}}{{#1\over#2}}{{#1\over#2}}}
\def\ket#1{\mathchoice{{%
\left|{#1}\right\rangle}}{|{#1}\rangle}{|{#1}\rangle}{|{#1}\rangle}}
\def\kettop#1{\left|{#1}\right\rangle_{\rm top}}
\def\ketSL#1{\left|{#1}\right\rangle_{\SL2}}
\def\ketGH#1{\left|{#1}\right\rangle_{\rm GH}}

\def\semi{\mathop{\rlap{\raisebox{1.5pt}{\tiny%
${\mid}$}}\kern-.5pt\mbox{\large$\times$}}}

\def\d{\partial}
\def\spsi{\psi^*}

\def\N#1{N\!=\!#1}

\def\SL#1{s\ell(#1)}
\def\tSL#1{{\widehat{s\ell}}(#1)}

\def\Jplus{J^+}
\def\Jminus{J^-}
\def\Jnaught{J^0}

\def\upleadsto{%
\begin{picture}(2,8)
\unitlength=1pt
\bezier{20}(0,-5)(-3,-2)(0,-1)
\bezier{20}(0,-1)(3,1)(0,3)
\bezier{20}(0,3)(-3,5)(0,7)
\bezier{10}(0,7)(1,8)(.5,9.5)
\put(1,11.3){\vector(0,1){2}}
\end{picture}%
}

\def\half{{\textstyle{1\over2}}}
\def\fourth{{\textstyle{1\over4}}}

\def\cA{{\cal A}}

\def\cO{{\cal O}}

\def\cU{{\cal U}}

\def\oN{{\open N}}
\def\oC{{\open C}}

\def\oZ{{\open Z}}

\def\ctop{{\ssf c}}
\def\Ctop{{\ssf C}}
\def\htop{{\ssf h}}

\def\hplus{{\ssf h}^+}
\def\hminus{{\ssf h}^-}
\def\jplus{{\ssf j}^+}
\def\jminus{{\ssf j}^-}
\def\theel{{\ssf l}}

\def\tensor{\otimes}

\def\const{{\rm const}}

\def\mM{{\ssf M}}
\def\mH{{\ssf H}}
\def\mR{{\ssf R}}
\def\mU{{\ssf U}}
\def\mV{{\ssf V}}

\def\smM{{\cal M}}

\def\smU{{\cal U}}
\def\smV{{\cal V}}
\def\smR{{\cal R}} 
\def\smW{{\cal W}} 

\def\CVER{{\cal CVER}}
\def\VER{{\cal VER}}
\def\CTVER{{\cal CTVER}}
\def\TVER{{\cal TVER}}
\def\CHW{{\cal CHW}}
\def\HW{{\cal HW}}
\def\CRHW{{\cal CRHW}}
\def\RHW{{\cal RHW}}

\def\TOP{{\cal TOP}}
\def\CMHW{{\cal CMHW}}
\def\MHW{{\cal MHW}}
\def\CRVER{{\cal CRVER}}
\def\RVER{{\cal RVER}}
\def\CMVER{{\cal CMVER}}
\def\MVER{{\cal MVER}}

\newtheorem{lemma}{Lemma}[section]

\newtheorem{thm}[lemma]{Theorem}

{\par\medskip}

{\noindent$\Box$\par\medskip} 

\def\emt{energy-momentum tensor}
\def\hw{highest-weight}

\def\NPB{Nucl.\ Phys.\ B}
\def\PLB{Phys.\ Lett.\ B}
\def\MPLA{Mod.\ Phys.\ Lett.\ A}

\def\IJMPA{Int.\ J.\ Mod.\ Phys.\ A}

\title{\hfill{\small\tt hep-th/9702074}\\
On the Equivalence of Affine $\SL2$ and $\N2$ Superconformal
Representation Theories\,\thanks{Contribution to the proceedings of the 30th
Int.\ Symposium Ahrenshoop on the theory of elementary particles, Buckow,
Germany, August 27--31, 1996.}}

\author{A.M.~Semikhatov\address{P.N.~Lebedev Physics Institute,
        53 Leninski prosp., Moscow 117924, Russia}%
         }

\begin{document}
\thispagestyle{empty}
\begin{abstract}
There exist two different languages, the $\tSL2$ and $\N2$ ones, to describe
similar structures; a dictionary is given translating the key
representation-theoretic terms related to the two algebras.  The main tool to
describe the structure of $\tSL2$ and $\N2$ modules is provided by
diagrams of extremal vectors. The $\tSL2$ and $\N2$ representation theories
of  certain \hw{} types turn out to be equivalent modulo the respective
spectral flows.
\end{abstract}

\maketitle

\section{INTRODUCTION}
In this talk I address a representation-theor\-etic problem that originates
in constructions as standard in conformal field theory as parafermions and
the Kazama--Suzuki (KS) construction, which relate the affine algebra $\tSL2$
and the $\N2$ superconformal (super-Lie)algebra in two dimensions.  The
results are conveniently expressed in the language of category theory.

Each of the two algebras has some relation to the bosonic string.  The matter
part of the string, which furnishes a representation of the Virasoro algebra,
is known to be described as the Hamiltonian reduction of $\tSL2$~\cite{[BO]}.
At the same time, dressing the matter theory into a non-critical bosonic
string gives rise to the $\N2$ superconformal algebra \cite{[GS2],[BLNW]}.
On the other hand, one should keep in mind that, as Table~\ref{tab:effluents}
shows, there is hardly anything that these two algebras appear to have in
common as regards their structure.  As regards the $\N2$ algebra, let me also
note that besides its appearance in the bosonic string (hence in all other
string theories, in that case as a subalgebra of larger superalgebras
\cite{[BLNW],[BLLS]}), it is the starting point in the construction of $\N2$
strings \cite{[Ade],[FT],[Marcus],[OV23],[Lechtenfeld],[S-sl21]}, which have
recently been suggested to play an important role in M-theory \cite{[KM]}.

\begin{table*}\small
\label{tab:effluents}
\renewcommand{\arraystretch}{0}
\begin{tabular}{|l|c|c|}
\hline\multicolumn{3}{|c|}{\rule{0pt}{1.5pt}}\\\hline
\strut algebra & $\tSL2$ & $\N2$\\
\hline\multicolumn{3}{|c|}{\rule{0pt}{1.5pt}}\\\hline
\strut {a Kac--Moody algebra?} & Yes & No \\\hline
\strut {contains fermions?} & No & Yes \\
\hline
\strut rank & 2 & 3 \\
\hline
\strut `Cartan' generators & $\mathstrut J^0_0,\,K$ & $L_0,\, H_0,\,\Ctop$\\
\hline
\strut `highest-weight' conditions&\hbox{$\mathstrut J^+_{\geq0}\!\approx\!
J^0_{\geq1}\!\approx\! J^-_{\geq1}\!\approx\!0$} &
\hbox{$L_{\geq1}\!\approx\! H_{\geq1}\!\approx\! Q_{\geq1}\!\approx\!
G_{\geq0}\!\approx\!0$
\phantom{${}_{\int_a^b}$}}\kern-13pt\\
\hline\multicolumn{3}{|c|}{\rule{0pt}{1.5pt}}\\\hline
\strut realized on the string worldsheet? & No & Yes \\
\hline\multicolumn{3}{|c|}{\rule{0pt}{1.5pt}}\\\hline
\end{tabular}
\end{table*}

In this talk, I describe some of the results of a work with Boris~Feigin and
Ilya~Tipunin~\cite{[FST]}, in which certain categories of representations of
the affine $\SL2$ and $\N2$ superconformal algebras are shown to be {\it
equivalent\/}.  The tools essential for the analysis of $\N2$ and $\tSL2$
modules include diagrams of extremal vectors and the spectral flow transform.
The idea to consider extremal vectors was put forward in \cite{[FS]}, where
it was observed that many representation-theoretic problems can be naturally
reformulated in terms of extremal vectors. An independent construction of
\cite{[ST3]} (and a similar one, \cite{[S-sl21sing]}) can be considered as a
manifestation of this general observation. I will illustrate in this talk
several basic points related to extremal diagrams, first of all how their
properties describe the structure of submodules of a given module (in
particular, its (sub)singular vectors). I consider such $\tSL2$ and $\N2$
modules that have isomorphic extremal diagrams, which results in the
equivalence of certain categories built out of these modules. This
requires introducing more general $\tSL2$ modules than those usually
considered.

The modules over the $\N2$ algebra that are generally~\cite{[BFK]} viewed as
`standard' Verma modules have an infinite number of equivalent \hw-like
states.  The affine $\SL2$ modules that correspond to these $\N2$ modules,
too, have an infinite number of `almost-\hw' states. They are called the
`relaxed' Verma modules, since they differ from the standard Verma modules by
somewhat `relaxed' \hw{} conditions. The ordinary $\tSL2$ Verma modules
constitute a class of submodules of relaxed Verma modules. Back to $\N2$, the
corresponding `smaller' class of modules are the so-called
`topological'~\cite{[ST2],[ST3]} (in fact, chiral) $\N2$ modules.  It thus
turns out that modules that look `standard' on the $\N2$ side correspond to
`less standard' $\tSL2$ modules, and vice versa, which is one of the reasons
for a `proliferation' of different types of modules I am going to deal with.
I will also have to {\it twist\/} (spectral-flow transform) both the $\tSL2$
and $\N2$ modules in order to compare \hw-type representation theories of the
two algebras, since the nature of the correspondence between the two
representation theories is such that it necessarily involves twisted modules
even if one starts with `untwisted' ones.

The main results are the pairwise equivalences of the respective corners of
the diagrams of categories of $\tSL2$ representations on the one hand and
$\N2$ representations on the other hand:
$$\!\!\new
\arraycolsep=2pt
\BA{ccc}
{\CHW}&\leadsto&{\CRHW}\\
\upleadsto&{}&\upleadsto\\
{\CVER}&\leadsto&{\CRVER}\\
{}&\kern-4pt{\tSL2}\kern-4pt&
\EA~{\rm and}~
\arraycolsep=2pt
\new\BA{ccc}
{\cal CTOP}&\leadsto&{\CMHW}\\
\upleadsto&{}&\upleadsto\\
{\CTVER}&\leadsto&{\CMVER}\\
{}&\kern-4pt{\N2}\kern-4pt&
\EA
$$
The categories are described as follows. `${\cal C}$' always stands for {\it
chains\/}, which I define below, of modules from the respective categories:
$\VER$ consists of the usual Verma modules over the affine $\SL2$ and all
their images under the spectral flow (i.e., the twisted Verma modules).
Category $\HW$ of twisted highest-weight-{\it type\/} modules is the
corresponding analogue of the $\cO$ category:  it is derived from $\VER$ by
taking all possible factor modules and `gluing' to each other different
modules (with the same twist); in what follows, I give an intrinsic
definition of this category using a criterion that is invariant under
twisting. Further, $\RVER$ are the relaxed Verma modules, which differ from
the usual Verma modules by one missing annihilation condition imposed on the
highest-weight vector and as a result possess infinitely many
`relaxed-highest-weight' vectors.  Then, $\RHW$ is the corresponding analogue
of the $\cO$ category of modules of the relaxed-highest-weight type.

On the $\N2$ side, $\TVER$ are {\it topological\/} Verma modules and all
their spectral flow transforms~\cite{[ST2],[ST3]}, while $\TOP$ are the
corresponding topological-highest-weight {\it type\/} modules.  The `massive'
Verma module category $\MVER$ consists of all possible twists of those
modules over the $\N2$ algebra that are commonly viewed as the ``standard"
$\N2$ Verma modules, while $\MHW$ is made up of modules of the same
highest-weight type, but not necessarily {\it Verma\/} modules.

Let me point out that, even though this is not indicated explicitly in the
names of the categories, each of the above categories includes {\it
twisted\/} modules (along with untwisted ones).

Taking chains of modules from these categories makes {\it the respective
corners of the two squares equivalent\/}, e.g.\ $\CVER\leadsto\CTVER$ and, at
the same time, $\CTVER\leadsto\CVER$, where the $\leadsto$ arrows are of
course {\it not\/} intertwining operators, nor any kind of morphisms of
modules, but rather, {\it functors\/}. Thus, any two modules related by a
morphism in one of the categories are $\leadsto$-mapped into modules related
by a morphism in the other category, and the claim of equivalence means in
particular that whatever properties a chosen morphism may have (embedding,
projection, \ldots), these are then preserved, while the composition of the
direct and the inverse arrows takes any object (a representation) into an
{\it isomorphic\/} object.

As long as submodules of Verma modules are associated with singular vectors,
a part of the statement amounts to the isomorphism between singular vectors
in the respective Verma modules. Singular vectors in topological $\N2$
modules allow a $1:1$ mapping into (well-known) singular vectors in $\tSL2$
Verma modules (as was claimed in~\cite{[S-sing]}); with the massive/relaxed
Verma modules, this is also true, but the structure of singular vectors is
more involved, and I will discuss it briefly for the $\tSL2$ case
(see~\cite{[FST]} for the details).  The point is that a given singular
vector does not necessarily generate a {\it maximal\/} submodule. This
situation can be described in terms of {\it subsingular\/} vectors:
constructing a system of sub-, subsub-, \ldots -singular vectors is nothing
but a way to describe the system of maximal submodules and of
(non-maximal) submodules thereof generated by vectors of a particular type
(those annihilated by a chosen set of operators from the algebra).

However, the structure of submodules can alternatively be described by
specifying those vectors that generate {\it maximal\/} submodules. I will
still call these vectors singular (and thus would no longer need the notion
of subsingular vectors), for the following reasons: for the algebras under
consideration, the vectors that generate maximal submodules turn out to
satisfy vanishing relations that are nothing but the spectral flow transform
of the `standard' annihilation conditions (i.e., of those imposed on \hw{}
vectors in untwisted modules).  Moreover, as I have already mentioned, the
relaxed/massive Verma modules can equally well be generated from an infinite
number of vectors each of which satisfies precisely the thus twisted \hw{}
conditions.  This makes it natural to consider \hw{} conditions up to the
spectral flow transform and to impose such \hw{} conditions on singular
vectors.  Among singular vectors understood in this broader sense, then, one
can always find those that generate maximal submodules, even though the
`standard' singular vector may generate a smaller submodule.

As we will see, it is advantageous to consider the entire {\it extremal
diagram\/} generated out of given \hw{} state and likewise, of a given
singular vector. It turns out that vectors that generate maximal submodules
-- which are singular in the broader sense that I adopt from now on -- and
the `standard' singular vectors belong to the same extremal diagram, and
moreover, it is the properties of the extremal diagram that are responsible
for whether or not the `standard' singular vector would generate a maximal
submodule.  In general, all the states in a given extremal diagrams satisfy
the same \hw{} conditions up to the spectral flow transform, however
stronger annihilation conditions may occur for some states in the diagram, in
which case the vectors in the extremal diagram may be divided into those
which do, and which do not, generate a maximal submodule.

The equivalence claim for {\it chains\/} of modules means that $\N2$ and
$\tSL2$ representation theories are equivalent modulo the respective spectral
flow transforms.

\section{The algebras}
\subsection{The affine $\SL2$ algebra}
The structure of $\tSL2$ Verma modules is conveniently encoded in
the {\it extremal diagram\/}
\BE
\unitlength=.8pt
\begin{picture}(220,80)
\put(35,0){
        \put(-32,62){\large $\ldots$}
        \put(0,60){$\bullet$}
        \put(15,65){${}^{J^-_0}$}
        \put(28,63){\vector(-1,0){22}}
        \put(30,60){$\bullet$}
        \put(45,65){${}^{J^-_0}$}
        \put(58,63){\vector(-1,0){22}}
        \put(60,60){$\bullet$}
        \put(75,65){${}^{J^-_0}$}
        \put(88,63){\vector(-1,0){22}}
        \put(90,60){$\circ$}
        \put(103,64){${}_{J^+_{-1}}$}
        \put(97,60){\vector(2,-1){17}}
        \put(115,47){$\bullet$}
        \put(24,-13){%
          \put(103,64){${}_{J^+_{-1}}$}
          \put(97,60){\vector(2,-1){17}}
          \put(115,47){$\bullet$}
        }
        \put(48,-26){%
          \put(103,64){${}_{J^+_{-1}}$}
          \put(97,60){\vector(2,-1){17}}
          \put(115,47){$\bullet$}
        }
        \put(72,-39){%
          \put(103,50){\large $\cdot$}
          \put(109,47){\large $\cdot$}
          \put(115,44){\large $\cdot$}
        }
}
\end{picture}
\label{Vermaextr}
\EE
which expresses the fact that $\Jplus_{-1}$ and $\Jminus_0$ are the
highest-level operators that do not yet annihilate the \hw{} state $\circ$.
All the other states in the module should be thought of as lying in the
interior of the wedge. In these conventions, e.g., $\Jnaught_{-1}$ is
represented as a downward vertical arrow. An important point is that
the diagram is angle-shaped, which reformulates as the following property:
take a state $\ket{v}$ represented by a point inside the angle and, for a fixed
$n\in\oZ$, consider all the states $(J^+_n)^i\,\ket{v}$, $i\in\oN$, and
$(J^-_{-n})^i\,\ket{v}$, $i\in\oN$.  These states fill out a straight line
in the diagram, which would {\it necessarily intersect the edge of the
diagram\/}, and therefore, in one of the directions, the action with either
$J^+_n$ or $J^-_{-n}$ terminates (vanishes) after a certain number of steps.

Automorphisms of the affine $\SL2$ algebra are the canonical involution and
the spectral flow
\BE
\cU_\theta:\new\BA{l}
J^+_n\mapsto J^+_{n+\theta},~
J^-_n\mapsto J^-_{n-\theta},\\
J^0_n\mapsto J^0_n+\frac{k}{2}\theta\delta_{n,0},
\EA\EE
where $\theta\in\oZ$, and $k$ is the level (I assume $k\neq-2$ in what
follows).  In general, Verma over the $\tSL2$ algebra are not invariant under
the spectral flow and are mapped into `twisted' modules.  A {\it twisted
Verma module\/} $\smM_{j,k,\theta}$ is freely generated by
$$
J^+_{\leq\theta-1},~
J^-_{\leq-\theta},~
J^+_{\leq-1}
$$
from a {\it ``twisted" highest-weight vector\/} $\ketSL{j,k;\theta}$ defined
by the conditions
\begin{eqnarray}
&&J^+_{\geq\theta}\,\ketSL{j,k;\theta}=
J^-_{\geq-\theta+1}\,\ketSL{j,k;\theta}=0,\nonumber\\
&&J^0_{\geq1}\,\ketSL{j,k;\theta}=0,\label{sl2higgeneral}\\
&&\left(J^0_{0}+\frac{k}{2}\theta\right)\,\ketSL{j,k;\theta}=
j\,\ketSL{j,k;\theta}\nonumber
\end{eqnarray}
Thus $\theta$ measures the gap between the mode numbers of $J^+$ and $J^-$
that annihilate the vacuum.
The respective extremal diagrams are `rotations' of \req{Vermaextr}.  A
crucial fact is that any straight line would still intersect the edge of the
diagram, and thus the argument discussed below~\req{Vermaextr} still applies.
I identify \hbox{$\ketSL{j,k}=\ketSL{j,k;0}$}, and denote
$\mM_{j,k}=\smM_{j,k;0}$.

All possible (integral) twists of Verma modules constitute the category
$\VER$.  However, already in the untwisted case, many
(if not all) interesting representations are {\it not} Verma modules, but
rather can be obtained from Verma modules by taking factors and `gluing'.
This gives the category $\cO$ (see \cite{[TheBook]}), in which the Verma
modules are `universal' objects in the sense that any irreducible
representation is a factor of a Verma module.  The standard definition of the
category $\cO$ singles out only the {\it untwisted\/} Verma modules
(those with $\theta=0$ in \req{sl2higgeneral}).  A remarkable fact is that
there exists an intrinsic definition of the category $\HW$ of highest-weight
{\it type\/} modules, which would include the twisted modules.  First of all,
to formalize the above observations, let $\ket{X}$ be an element of a module
over the affine $\SL2$ algebra and let us fix an integer $\theta$. For $J$
being either $\Jplus$ or $\Jminus$, we say that the $J_\theta$-chain
terminates on $\ket{X}$, and write \ $(J_\theta)^{+\infty}\,\ket{X}=0$, if
$$
\exists N\in\oZ,~ n\geq N :
(J_\theta)^n\,\ket{X}=0\,.
$$
Further, all the modules in what follows {\it are assumed graded with respect
to the Cartan subalgebra of the respective algebra\/}.  I will use the
criterion of terminating chains to define categories of \hw-{\it type\/}
representations. This and similar criteria will be applied to $\Jplus$ and
$\Jminus$ generators in the $\tSL2$ case and to $G$ and $Q$ in the $\N2$
case. For brevity, I will give explicitly only those parts of definitions
that have to do with twisting, omitting explicit stipulations of the standard
$\cO$-category requirements with respect to the remaining generators
($\Jnaught$, and $L$ and $H$ respectively), which state that {\it acting with
the annihilation operators spans out a finite-dimensional space\/}.  Then, an
$\tSL2$  module $\smU$ belongs to the {\it category $\HW$ of
$\tSL2$ twisted \hw-type representations\/} if, for any element $\ket{X}$ of
$\smU$, $\forall n\in\oZ$
\BE
\BA{ll}
{\rm either}&(J^+_{n})^{+\infty}\,\ket X=0\\
{\rm or}    &(J^-_{-n})^{+\infty}\,\ket X=0
\EA\label{Jterminate}
\EE
(and, in accordance with the above remarks, it is tacitly assumed that $\cU$
is graded and that acting with $(\Jnaught_m)^n$, $m,n\geq1$, on any vector
produces a finite-dimensional space).

Singular vectors in $\tSL2$ Verma modules $\mM_{j,k}$
are defined in the standard way. To explicitly
construct singular vectors, one introduces the objects $(J^+_{-1})^\alpha$
and $(J^-_0)^\alpha$ that implement the action of generators of the
affine Weyl group on the space of highest-weights, see \cite{[MFF]} for the
details.
These objects correspond to reflections with respect to two
positive simple roots of the affine $\SL2$ algebra.
In fact $(J^+_{-1})^\alpha$ and $(J^-_0)^\alpha$ define the following Weyl
group action on the line $k=\const$ in the $kj$ plane of highest-weights:
\BE\new\BA{l}
(J^-_0)^{2j+1}\,:\,\ketSL{j,k}\rightarrow\ketSL{-1-j,k},\kern-20pt\\
(J^+_{-1})^{k+1-2j}\,:\,\ketSL{j,k}\rightarrow\ketSL{k+1-j,k}\,.\kern-20pt
\label{sl2weylaction}\EA\EE
The action of $(J^+_{-1})^\alpha$ and $(J^-_0)^\alpha$ can be extended
from the set of highest-weight vectors to the Verma modules over these
vectors.
Then,
\begin{thm}\mbox{}\nopagebreak

{\rm I. (\cite{[KK]})}
A singular vector exists in the module $\mM_{j,k}$ iff $j=\jplus(r,s,k)$ or
$j=\jminus(r,s,k)$, where
\BE\left.\new\BA{l}
\jplus(r,s,k)=\frac{r-1}{2}-(k+2)\frac{s-1}{2}\\
\jminus(r,s,k)=-\frac{r+1}{2}+(k+2)\frac{s}{2}
\EA\kern-4pt\right\}\BA{l}r,s\in\oZ\\
k\in\oC\EA
\label{sl2singcond}\EE

{\rm II. (\cite{[MFF]})}
All singular vectors $\ket{S_{\pm}^{\rm MFF}(r,s,k)}$ in the Verma module
$\mM_{j,k}$ over the affine $\SL2$ algebra are given by the explicit
construction:
\BE
\new\BA{l}
\ket{S_+^{\rm MFF}(r,s,k)}={}\\
\quad(J^-_0)^{r+(s-1)(k+2)}(J^+_{-1})^{r+(s-2)(k+2)}\ldots\\
\qquad{}\cdot(J^+_{-1})^{r-(s-2)(k+2)}(J^-_0)^{r-(s-1)(k+2)}\kern-20pt\\
\qquad\quad{}\cdot\ket{\jplus(r,s,k),k}_{\SL2},\\
\ket{S_-^{\rm MFF}(r,s,k)}={}\\
\quad(J^+_{-1})^{r+(s-1)(k+2)}(J^-_0)^{r+(s-2)(k+2)}\ldots\\
\qquad{}\cdot(J^-_0)^{r-(s-2)(k+2)}(J^+_{-1})^{r-(s-1)(k+2)}\kern-20pt\\
\qquad\quad{}\cdot\ket{\jminus(r,s,k),k}_{\SL2}\,.
\EA
\label{mffminus}
\label{mffplus}
\EE
\end{thm}
Singular vectors in twisted Verma modules follow by
applying the spectral flow transform to~\req{mffplus}.

A more general class of affine $\SL2$ modules can be introduced by relaxing
the annihilation conditions \req{sl2higgeneral}:
For $\theta\in\oZ$, a {\it relaxed twisted Verma module\/}
$\smR_{j,\Lambda,k;\theta}$ is freely generated by the operators
\ $J^+_{\leq\theta}$, $J^-_{\leq-\theta}$, and $J^0_{\leq-1}$ \
from the state $\ketSL{j,\Lambda,k;\theta}$ that satisfies the annihilation
conditions
\begin{eqnarray}
&&\kern-16pt J^+_{\geq\theta+1}\,\ketSL{j,\Lambda,k;\theta}=J^0_{\geq1}\,
\ketSL{j,\Lambda,k;\theta}={}\nonumber\\
&&\kern-16pt J^-_{\geq-\theta+1}\,\ketSL{j,\Lambda,k;\theta}=0\,.
\label{floorhw}
\end{eqnarray}
and
\begin{eqnarray}
&&\kern-16pt\left(J^0_0+\frac{k}{2}\theta\right)\,
\ketSL{j,\Lambda,k;\theta}=j\,\ketSL{j,\Lambda,k;\theta},\nonumber\\
&&\kern-16pt
\left(J^-_{-\theta}J^+_\theta+(k+2)\theta(j-\frac{k}{4}\theta)\right)\,
\ketSL{j,\Lambda,k;\theta}={}\nonumber\\
&&\kern-16pt\qquad\qquad\Lambda\,\ketSL{j,\Lambda,k;\theta}\,.
\end{eqnarray}

The corresponding extremal diagram opens up to the straight  angle; in the
untwisted case $\theta=0$ it thus becomes
\BE
\unitlength=.8pt
\begin{picture}(200,20)
\put(20,0){%
        \put(-22,2){$\ldots$}
        \put(0,0){$\bullet$}
        \put(15,5){${}^{J^-_0}$}
        \put(28,3){\vector(-1,0){22}}
        \put(30,0){$\bullet$}
        \put(45,5){${}^{J^-_0}$}
        \put(58,3){\vector(-1,0){22}}
        \put(60,0){$\bullet$}
        \put(75,5){${}^{J^-_0}$}
        \put(88,3){\vector(-1,0){22}}
        %
        \put(90,0){$\star$}
        \put(100,5){${}^{J^+_0}$}
        \put(97,3){\vector(1,0){22}}
        \put(120,0){$\bullet$}
        \put(130,5){${}^{J^+_0}$}
        \put(127,3){\vector(1,0){22}}
        \put(150,0){$\bullet$}
        \put(160,5){${}^{J^+_0}$}
        \put(157,3){\vector(1,0){22}}
        \put(180,0){$\bullet$}
        \put(193,2){$\ldots$}
}
\end{picture}
\label{floor}
\EE
The state marked with $\star$ is the above $\ketSL{j,\Lambda,k;\theta}$.
The other states $\ketSL{j,\Lambda,k;\theta|n}$, $n\in\oZ$, from
the extremal diagram are
$$
\!\ketSL{j,\Lambda,k;\theta|n}\!=\!\left\{\kern-6pt\new\BA{l}
(J^-_{-\theta})^{-n}\ketSL{j,\Lambda,k;\theta},n<0,\\
(J^+_\theta)^{n}\,\ketSL{j,\Lambda,k;\theta},~n>0,
\EA\right.
$$
with $\ketSL{j,\Lambda,k;\theta|0}=\ketSL{j,\Lambda,k;\theta}$.
I also define $\ketSL{j,\Lambda,k|n}=\ketSL{j,\Lambda,k;0|n}$.

In the generic case, one can travel both ways along the extremal diagram:
for example, the `untwisted' diagram \req{floor} ($\theta=0$) acquires a
`fat' form
\BE
\unitlength=.8pt
\begin{picture}(200,30)
\put(30,0){%
        \put(-21,12){$\ldots$}
        \put(0,10){$\bullet$}
        \put(15,17){${}^{J^-_0}$}
        \put(28,15){\vector(-1,0){22}}
        \put(7,11){\vector(1,0){22}}
        \put(15,3){${}_{J^+_0}$}
        \put(30,10){$\bullet$}
        \put(45,17){${}^{J^-_0}$}
        \put(58,15){\vector(-1,0){22}}
        \put(37,11){\vector(1,0){22}}
        \put(45,3){${}_{J^+_0}$}
        \put(60,10){$\bullet$}
        \put(75,17){${}^{J^-_0}$}
        \put(88,15){\vector(-1,0){22}}
        \put(67,11){\vector(1,0){22}}
        \put(75,3){${}_{J^+_0}$}
        \put(90,10){$\star$}
        \put(105,17){${}^{J^-_0}$}
        \put(118,15){\vector(-1,0){22}}
        \put(97,11){\vector(1,0){22}}
        \put(105,3){${}_{J^+_0}$}
        \put(120,10){$\bullet$}
        \put(135,17){${}^{J^-_0}$}
        \put(148,15){\vector(-1,0){22}}
        \put(127,11){\vector(1,0){22}}
        \put(135,3){${}_{J^+_0}$}
        \put(150,10){$\bullet$}
        \put(165,17){${}^{J^-_0}$}
        \put(178,15){\vector(-1,0){22}}
        \put(157,11){\vector(1,0){22}}
        \put(165,3){${}_{J^+_0}$}
        \put(180,10){$\bullet$}
        \put(193,12){$\ldots$}
}
\end{picture}
\label{bothways}\EE
where the
composition of the direct and the inverse arrows results in each case
only in a factor:
$$\new
\arraycolsep=1pt
\BA{rcl}
\underline{n\leq0\,{:}\mathstrut}\hfill\\
  J^-_0\,\ketSL{j,\Lambda,k|n}&=&\ketSL{j,\Lambda,k|n-1},\\
{}J^+_0\ketSL{j,\Lambda,k|n\!-\!1}&=&(\Lambda\!-\!n(n\!-\!1)\!-\!2(n\!-\!1)j)
\cdot\\
    {}&{}&\quad{}\cdot\ketSL{j,\Lambda,k|n},\\
\underline{n\geq0\,{:}\mathstrut}\hfill\\
  J^+_0\,\ketSL{j,\Lambda,k|n}&=&\ketSL{j,\Lambda,k|n+1},\\
{}J^-_0\,\ketSL{j,\Lambda,k|n+1}&=&(\Lambda-n(n+1)-2nj)\cdot\\
    {}&{}&\quad{}\cdot\ketSL{j,\Lambda,k|n}
\EA$$
However, this factor may vanish for some values of the parameters and this
gives rise to standard {\it Verma\/} submodules.  Thus, whenever the
parameters are such that, e.\,g., $J^+_0\approx0$ at a certain step, {\it
properties of the extremal diagram change\/}, and one cannot come back to the
$\star$ state by acting with $J^+_0$:
\begin{eqnarray}
&&{n\leq-1,\,\mathstrut}
\Lambda=n(n+1)+2nj~\Longrightarrow\qquad\qquad\nonumber\\
&&\qquad\qquad J^+_0\ketSL{j,\Lambda,k|n}=0\,.\label{Vermaneg}
\end{eqnarray}
One keeps on acting with $J^+_{-1}$ instead (one mode down), and
thus the extremal diagram becomes
\BE
\unitlength=.72pt
\begin{picture}(200,62)
\put(29,-13){%
        \put(-25,62){\Large $\ldots$}
        \put(0,60){$\bullet$}
        \put(15,67){${}^{J^-_0}$}
        \put(28,65){\vector(-1,0){22}}
        \put(7,61){\vector(1,0){22}}
        \put(15,55){${}_{J^+_0}$}
        \put(30,60){$\bullet$}
        \put(45,67){${}^{J^-_0}$}
        \put(58,65){\vector(-1,0){22}}
        \put(37,61){\vector(1,0){22}}
        \put(45,55){${}_{J^+_0}$}
        \put(60,60){$\bullet$}
        \put(75,67){${}^{J^-_0}$}
        \put(88,65){\vector(-1,0){22}}
        \put(67,61){\vector(1,0){22}}
        \put(75,55){${}_{J^+_0}$}
        \put(90,60){$\circ$}
        \put(105,67){${}^{J^-_0}$}
        \put(118,65){\vector(-1,0){22}}
        \put(120,60){\raisebox{1.5pt}{${\scriptscriptstyle\odot}$}}
        \put(135,67){${}^{J^-_0}$}
        \put(148,65){\vector(-1,0){22}}
        \put(127,61){\vector(1,0){22}}
        \put(137,55){${}_{J^+_0}$}
        \put(150,60){$\bullet$}
        \put(40,0){
        \put(150,60){$\bullet$}
         \put(165,67){${}^{J^-_0}$}
         \put(178,65){\vector(-1,0){22}}
         \put(157,61){\vector(1,0){22}}
         \put(165,55){${}_{J^+_0}$}
         \put(180,60){$\star$}
        }
        \put(163,62){\Large$\ldots$}
        \put(230,62){\Large$\ldots$}
        \put(90,38){${}^{J^+_{-1}}$}
        \put(95,58){\vector(2,-1){17}}
        \put(114,53){\vector(-2,1){17}}
        \put(115,47){$\bullet$}
        \put(24,-13){%
        \put(90,38){${}^{J^+_{-1}}$}
        \put(95,58){\vector(2,-1){17}}
        \put(101,57.5){${}^{J^-_{1}}$}
        \put(114,53){\vector(-2,1){17}}
        \put(115,47){$\bullet$}
        }
        \put(48,-26){%
        \put(90,38){${}^{J^+_{-1}}$}
        \put(95,58){\vector(2,-1){17}}
        \put(106,55){${}^{J^-_{1}}$}
        \put(114,53){\vector(-2,1){17}}
        \put(115,47){$\bullet$}
        }
        \put(72,-39){%
          \put(103,50){\Large $\cdot$}
          \put(109,47){\Large $\cdot$}
          \put(115,44){\Large $\cdot$}
        }
}
\end{picture}
\label{withVerma}\EE
where in the subdiagram one recognizes (the fat form of) the extremal
diagram~\req{Vermaextr}. Therefore, any Verma module can be thought of as a
{\it submodule\/} of a relaxed Verma module.

Similarly, one may have $J^-_0\approx0$ at a certain stage in the
diagram \req{bothways},
\begin{eqnarray}
&&{n\geq1{,}\quad\mathstrut}
\Lambda=n(n-1)+2(n-1)j~\Longrightarrow\nonumber\\
&&\qquad\qquad{}J^-_0\ketSL{j,\Lambda,k|n}=0\,.\label{Vermapos}
\end{eqnarray}
Then the branching of the extremal diagram is a mirror image
of \req{withVerma}, and the Verma submodule is given by the spectral flow
transform with $\theta=1$ of a standard Verma module.

By a mere application of the spectral flow, the above results reformulate for
{\it twisted\/} relaxed Verma modules.

\medskip

As in the standard Verma case, the category $\RVER$ of all twisted relaxed
Verma modules can be extended to a larger category $\RHW$ of arbitrary
(twisted) relaxed-highest-weight {\it type\/} modules.  Now that the angle in
the extremal diagrams has opened up to the straight angle, there would
certainly exist in the extremal diagrams straight lines infinite on both
sides.  However, a condition on the class of modules can still be given in
the form of the requirement that, starting with any vector from a given
module, the action with any of the {\it bent $J^\pm$-chains\/}
$$
\unitlength=1pt
\begin{picture}(130,10)
\put(-30,0){
        \put(100,-2.5){$\cdot$}
        \put(105,1.5){\vector(4,1){20}}
        \put(98,2){\vector(-2,1){17}}
}
\end{picture}
$$
terminates:
A module $\smU$ over the affine $\SL2$ algebra is said to belong to the {\it
category $\RHW$ of twisted relaxed $\tSL2$ \hw\ representations\/} if, for
any element $\ket{X}$ of $\smU$, $\forall \theta\in\oZ$
\BE
\BA{ll}
{\rm either}&(J^+_{\theta})^{+\infty}\,\ket X=0\\
{\rm or}    &(J^-_{-\theta+1})^{+\infty}\,\ket X=0
\EA
\EE
Again, this condition is invariant under the spectral flow, and thus selects
all the twisted modules, none of which would be `overrelaxed' in the sense of
its extremal diagram occupying {\it more\/} than half a plane.

I will now briefly describe singular vectors in relaxed Verma modules
with $\theta=0$ (similar results for modules with $\theta\neq0$ can be
obtained immediately by applying the spectral flow transform).  These
singular vectors are naturally split into two groups: those occurring at the
edge of the extremal diagram, and those occurring `inside' the extremal
diagram. As I have shown, the former immediately follow from the analysis
of extremal diagrams.  It therefore remains to consider the second type of
singular vectors; they do not signify the presence of a Verma submodule, but
rather represent a {\it relaxed Verma submodule\/}, i.e., another straight
line in the extremal diagram parallel to the edge of the diagram, with every
state in the subdiagram satisfying the relaxed \hw{} conditions\req{floorhw};
interesting things start to happen when some of these states satisfy
stronger, {\it Verma\/}, \hw{} conditions.

Given a state $\ket{j, \Lambda, k}$, consider $(\Jminus_0)^{-\mu}\,\ket{j,
\Lambda, k}$ with $\mu = \jminus(r, s, k) - j$. Whenever $\Lambda$ is chosen
as $\Lambda(r, s, j, k)$ where
$$\kern-6pt\new\BA{l}
\Lambda(r, s, j, k)={}\\
{}~\fourth(-1 - 2 j - r + 2 s + k s) (1 + 2 j - r + 2 s + k s),
\EA$$
the `continued' state $(\Jminus_0)^{-\mu}\,\ketSL{j, \Lambda, k}$ would
satisfy the {\it Verma\/} \hw{} conditions, with the spin given by
$\jminus(r,s,k)$. Therefore the usual ${\rm MFF}^-$ singular vector can be
constructed on this state, as
$$
\!{\cal MFF}^-(r, s, k)(\Jminus_0)^{j-\jminus(r, s, k)}\!
\ketSL{j, \Lambda(r, s, j, k),k}
\label{work10}$$
where ${\cal MFF}^-$ is the singular vector {\it operator\/} (read off by
dropping the \hw{} state in~\req{mffminus}). This has to be mapped back to
the original relaxed Verma module. In particular, no non-integral powers of
$\Jminus_0$ should remain, which is achieved by acting on \req{work10} with
$(\Jminus_0)^{\jminus(r, s, k) - j + N}$, where $N$ is an integer. However,
to be left after the rearrangements with only {\it positive\/} integral
powers, the integer $N$ has to be $\geq r+rs$. I thus choose
$$\new\BA{l}
\Sigma^-(r,s,j,k)={}\\
\quad(\Jminus_0)^{\jminus(r, s, k) - j + r + r s}\,
{\cal MFF}^-(r, s, k)\cdot{}\\
\qquad{}\cdot(\Jminus_0)^{j-\jminus(r, s, k)}\,
\ketSL{j, \Lambda(r, s, j, k), k}
\EA$$
as a representative of the singular vector in the relaxed Verma
module~$\smR_{j,\Lambda(r, s, j, k),k;0}$. The rules for dealing with
non-integral powers are directly analogous to those used in the standard,
Verma, MFF construction.

Similarly, for $\mu=\jplus(r, s + 1, k) - j$, the state $(\Jplus_0)^{1 +
\jplus(r, s + 1, k) - j}\,\ketSL{j, \Lambda(r, s, j, k), k}$ is formally a
Verma \hw{} state twisted by the spectral flow transform with $\theta=1$, and
thus the singular vector becomes
$$\new\BA{l}
\Sigma^+(r,s,j,k)={}\\
\quad(\Jplus_0)^{j + r + r s-1 - \jplus(r, s + 1, k)}\,
{\cal MFF}^{+, 1}(r, s, k)\cdot{}\\
\qquad{}\cdot(\Jplus_0)^{1 + \jplus(r, s + 1, k) - j}\,\ketSL{j, \Lambda(r, s, j, k), k}
\EA
$$

These singular vectors can be acted upon with $J^\pm_0$, which may allow one
to map $\Sigma^+$ and $\Sigma^-$ to the same grade. Whenever this is
possible, the two vectors mapped to the same grade are linearly dependent,
and thus there exists a unique relaxed singular vector (in terms of extremal
diagrams, $\Sigma^+$ and $\Sigma^-$ then generate the same extremal
subdiagram).  However, the action of $J^+_0$ or $J^-_0$ may give zero at some
step, which would mean encountering {\it a Verma \hw{} state\/} in the
subdiagram representing the singular vector. In that case, it may still be
possible to act with $(J^-_0)^{-1}$ or $(J^+_0)^{-1}$ respectively, the
latter being understood as one of the `continued' operators from
\req{sl2weylaction}.  The extremal diagram becomes, schematically,
$$
\unitlength=.8pt
\begin{picture}(250,110)
\put(0,20){
{
 \put(138,85){\vector(1,0){110}}
 \put(129,85){\vector(-1,0){120}}
 \put(40,85){\vector(3,-2){160}}
 \put(72,0){\vector(-1,0){70}}
 \put(72,0){\vector(3,-2){35}} 
 \put(219,0){$\vector(-1,0){146}$} 
\put(226,0){$\vector(1,0){30}$}
}
\put(130,83){$\star$}
\put(30,1){${}^{\Sigma^-}$}
\put(30,-3){$\bullet$}
\put(46,10){${}_{{\rm MFF}^-}$}
\put(220,1){${}^{\Sigma^+}$}
\put(220,-3){$\bullet$}
%
%
\linethickness{.7pt}
\bezier{12}(70,0)(83,16)(94,0)
\put(91.5,4){\vector(1,-2){2}}
\put(78,17){${}_{(J^-_0)^{-1}}$}
\linethickness{1.2pt}
\put(219,0){$\vector(-1,0){125}$} 
}
\end{picture}
$$
where $\Sigma^-$ happens to lie inside the Verma submodule built on
the usual MFF singular vector; the latter is {\it necessarily\/}
embedded into a {\it Verma\/} submodule of the type of the one pictured
in~\req{withVerma}.

A similar `slope' may further be encountered when moving on the left from
$\Sigma^+$, by acting on it with powers of $J^-_0$. Whenever this happens, a
further {\it Verma\/} \hw{} state would appear in the top floor of the
diagram. Yet it may still be possible to generate the entire lower floor from
$\Sigma^+$ by acting with $(J^+_0)^{-1}$ where the action of $J^-_0$
vanishes.

By a generalized $J^+_0$-descendant {\rm(}respectively, generalized
$J^-_0$-descendant{\rm)} of a state $\ket{v}$ in the relaxed Verma module
${\ssf R}$, I will mean any state that can
be obtained from $\ket{v}$ by an arbitrary number of the following steps
i)~acting with $J^+_0$ {\rm(}resp.  $J^-_0${\rm)} whenever the result is
non-vanishing, and \ ii)~acting with $(J^-_0)^{-1}$ {\rm(}resp.
$(J^+_0)^{-1}${\rm)} whenever the result is defined as an element in ${\ssf
R}$ and step~i) cannot be applied.
\begin{thm}\mbox{}\nopagebreak

{\rm I.}~In the general position, singular vectors $\Sigma^+(r,s,j,k)$ and
$\Sigma^-(r,s,j,k)$ are different representatives for the same singular
vector in the relaxed Verma module $\mR_{j,\Lambda(r,s,j,k),k}$: the
generalized $J^\pm_0$-descendants of $\Sigma^+(r,s,j,k)$ and
$\Sigma^-(r,s,j,k)$ that are in the same grade are proportional to each
other.

{\rm II.}~Whenever
$$
j=-\half(1 + m + n)\,,~
k+2=\frac{-m + n + r}{s}\,,
$$
with $n\in-\oN$ and $m\in\oN$, there exist generalized $J^\pm_0$-descendants
of $\Sigma^+$ and $\Sigma^-$ in the same grade that are linearly independent.
\end{thm}
In the second case, there thus exist two linearly independent singular
vectors in the same grade.  These linearly independent singular vectors then
reside in a section of the length $2r - m + n$ of the lower floor of the
extremal diagram.  It follows that, whenever two different singular vectors
in the same grade appear, these are necessarily the usual MFF singular
vectors in the corresponding Verma submodules.
$$
\unitlength=.8pt
\begin{picture}(250,120)
\put(-10,20){
{
 \put(138,85){\vector(1,0){130}}
 \put(129,85){\vector(-1,0){120}}
 \put(90,85){\vector(3,-2){153}}  
 \put(210,85){\vector(-3,-2){153}}  
 \put(175,-.6){\vector(3,-2){37}} 
 \put(130,.6){\vector(-3,-2){37}} 
 \put(130,.6){\vector(1,0){135}} 
 \put(175,-.6){\vector(-1,0){145}} 
 \put(193,.6){\vector(-3,-2){37}} 
 \put(110,-.6){\vector(3,-2){35}} 
}
\put(130,83){$\star$}
\put(65,1){${}^{\Sigma^-}$}
\put(65,-3.6){$\bullet$}
\put(225,1){${}^{\Sigma^+}$}
\put(225,-2.4){$\bullet$}
\put(163,8){${}_{\ket{{\rm mff}}^-}$}
\put(118,8){${}_{\ket{{\rm mff}}^+}$}
         \linethickness{.2pt}
\put(130,90){\vector(-1,0){40}}
\put(130,90){\vector(1,0){2}}
\put(110,92){${}^{-n}$}
\put(131,-5){\vector(1,0){45}}
\put(131,-5){\vector(-1,0){2}}
\put(132,-17){${}^{2r-m+n}$}
\put(137,90){\vector(1,0){72}}
\put(137,90){\vector(-1,0){2}}
\put(166,92){${}^{m+1}$}
}
\end{picture}
$$
I refer to \cite{[FST]} for more pictures and a detailed description of all
the possible cases.

\subsection{Auxiliaries}
I will need a fermionic system ($bc$ ghosts),
defined in terms of operator products as $B(z)\,C(w)={1\over z-w}$
with the \emt\ $T^{\rm GH}=-B\,\d C$.
Denote by $\Omega$ the module generated from the vacuum $\ketGH0$
defined by the conditions
\BE
C_{\geq1}\ketGH0=B_{\geq0}\ketGH0=0
\EE
The thus defined vacuum is an $sl_2$-invariant state \cite{[FMS]}; one  can
choose other highest-weight states, $\ketGH\lambda$, which belong to the same
module and are determined by
\BE
C_{\geq1-\lambda}\ketGH\lambda=B_{\geq\lambda}\ketGH\lambda=0\,.
\label{lambdavac}
\EE

The states $\ketGH\lambda$ with different $\lambda$ can be connected by means
of operators $c(\mu,\nu)$ and $b(\mu,\nu)$ which are products of
fermionic modes
$$
c(\mu,\nu)=\prod_{n=1}^{\nu-\mu+1}C_{\mu+n},\quad
b(\mu,\nu)=\prod_{n=1}^{\nu-\mu+1}B_{\nu+n},
$$
with $\nu-\mu+1\in\oN$, and which map a vector $\ketGH\lambda$ as follows
$$\new\BA{l}
c(-\lambda-\ell+1,-\lambda)\,:\,\ketGH\lambda\mapsto
\ketGH{\lambda+\ell},\\
b(-\lambda-\ell,\lambda-1)\,:\,\ketGH\lambda\mapsto
\ketGH{\lambda-\ell},
\EA\ell\in\oN
$$

Let me now introduce a ``Liouville" scalar which will be used to `invert' the
KS mapping.  This is just a free scalar, called `Liouville' for its
signature, \ $\phi(z)\phi(w)=-\ln(z-w)$.  I define vertex operators, referred
to as `antifermions', $\psi=e^\phi$ and $\spsi=e^{-\phi}$.
The energy-momentum tensor is taken to be
\BE
T_\phi=-\half\d\phi\d\phi+\half\d^2\phi\,.
\label{pseudoemt}\EE

\subsection{$\N2$\label{subsec:N2}}
Nonvanishing commutation relations of the $\N2$ superconformal algebra
$\cA$ can be chosen as 
\BE\new
\arraycolsep=1pt
\kern-2pt\BA{rcllcl}
{[}L_m,L_n]&=&(m\!-\!n)L_{m+n},&[ H_m, H_n]&=
&\frac{\Ctop}{3}m\delta_{m+n},\kern-30pt\\
{[}L_m, G_n]&=&(m\!-\!n) G_{m+n},&[ H_m, G_n]&=& G_{m+n},\kern-30pt\\
{[}L_m, Q_n]&=&-n Q_{m+n},&[ H_m, Q_n]&=&- Q_{m+n},\kern-30pt\\
{[}L_m, H_n]&=&\multicolumn{4}{l}{-n H_{m+n}+\frac{\Ctop}{6}(m^2+m)
\delta_{m+n},\kern-30pt}\\
\{G_m, Q_n\}&=&\multicolumn{4}{l}{2L_{m+n}\!-\!2n H_{m+n}\!+\!
\frac{\Ctop}{3}(m^2\!+\!m)\delta_{m+n},\kern-30pt}
\\
\multicolumn{6}{r}{m, n\in\oZ.\kern-30pt}
\EA\label{topalgebra}
\EE

When applied to the generators of \req{topalgebra}, the spectral flow
transform~\cite{[SS],[LVW]} $\cU_\theta$ acts as
\BE\new\BA{rcl}
L_n&\mapsto&L_n+\theta H_n+\frac{\Ctop}{6}(\theta^2+\theta)
\delta_{n,0},\\
 H_n&\mapsto& H_n+\frac{\Ctop}{3}\theta\delta_{n,0},\\
 Q_n&\mapsto& Q_{n-\theta},\quad  G_n\mapsto G_{n+\theta}\,
\EA\label{U}\EE
This gives the algebra $\cA_\theta$, which is isomorphic to the $\N2$
superconformal algebra and whose generators $L^\theta_n$, $ Q^\theta_n$,
$H^\theta_n$ and $ G^\theta_n$ can be taken as the RHSs of \req{U}.
One thus obtains the Neveu--Schwarz and Ramond $\N2$ algebras, as well as the
algebras in which the fermion modes range over $\pm\theta+\oZ$,
$\theta\in\oC$.

Now I define twisted topological\,\footnote{The name is inherited from the
non-critical bosonic string, where matter vertices can be dressed into $\N2$
primaries that satisfy the highest-weight conditions \req{gentophwint}; in
that context, the algebra \req{topalgebra} is viewed as a topological
algebra.} Verma modules $\smV_{h,t;\theta}$ over the $\N2$ algebra. This is
the module generated from the {\it topological highest-weight vector\/}
$\ket{h,t;\theta}_{\rm top}$ defined by
\BE\kern-3pt
\arraycolsep=2pt\new\BA{rcll}
L_m\ket{h,t;\theta}_{\rm top}&=&0,&H_m\ket{h,t;\theta}_{\rm top}=0\,,\ m\geq1,
\kern-60pt\\
 Q_\lambda\ket{h,t;\theta}_{\rm top}&=&0,&\lambda\in-\theta+\oN_0\,,\\
 G_\nu\ket{h,t;\theta}_{\rm top}&=&0,&\nu\in\theta+\oN_0
\EA~\quad\theta\in\oZ\,;
\label{gentophwint}
\EE
and for the `Cartan' generators,
\BE\new
\BA{l}
( H_0+\frac{\Ctop}{3}\theta)\,\kettop{h,t;\theta}=
h\,\kettop{h,t;\theta},\\
(L_0+\theta H_0+\frac{\Ctop}{6}(\theta^2+\theta))
\,\kettop{h,t;\theta}=0\,\\
\Ctop\,\ket{h,t;\theta}_{\rm top}=\frac{3(t-2)}{t}\,\ket{h,t;\theta}_{\rm top}
\EA
\label{genhw}\EE
The $\theta=0$ case describes the `{\it ordinary\/}' topological Verma
modules $\mV_{h, t}\equiv\smV_{h,t;0}$, with the corresponding topological
\hw{} vector $\kettop{h,t}\equiv\kettop{h,t;0}$.

The extremal diagram of a topological Verma module reads
\BE
\unitlength=.9mm
\begin{picture}(70,40)
\put(25,5){
        \put(00.00,00.00){$\bullet$}
        \put(10.00,20.00){$\bullet$}
        \put(10.00,20.00){$\bullet$}
        \put(20.00,30.00){$\bullet$}
        \put(29.70,20.00){$\bullet$}
        \put(40.00,00.00){$\bullet$}
        \put(9.70,19.00){\vector(-1,-2){8}}
        \put(19.70,29.70){\vector(-1,-1){7}}
        \put(22.00,29.70){\vector(1,-1){7}}
        \put(32.00,19.00){\vector(1,-2){8}}
        \put(00.00,13.00){${}_{G_{-2}}$}
        \put(11.00,28.00){${}_{G_{-1}}$}
        \put(27.00,28.00){${}_{Q_{-1}}$}
        \put(37.00,13.00){${}_{Q_{-2}}$}
        \put(-17.00,00.00){${}^{\ket{h+2,t;-2}}$}
        \put(-7.00,21.00){${}_{\ket{h+1,t;-1}}$}
        \put(19.00,34.00){${}_{\ket{h,t}_{\rm top}}$}
        \put(33.00,21.00){${}_{\ket{h-1,t;2}}$}
        \put(43.00,00.00){${}^{\ket{h-2,t;3}}$}
        \put(00.50,-06.00){$\vdots$}
        \put(40.50,-06.00){$\vdots$}
        }
\end{picture}
\label{newdiagram2}\EE
An important point here is the existence of a `cusp', i.e. a state that
satisfies stronger \hw{} than the other states in the diagram.

Next, I need the concept of terminating fermionic chains.  Let $F$ denote
either $Q$ or $G$, and $\ket{X}$ be an element of a module over the $\N2$
algebra. Fix also an integer $n$. We say that the fermionic $F$-chain
terminates on $\ket{X}$, and write
$\ldots\,F_{n-3}\,F_{n-2}\,F_{n-1}\,F_{n}\,\ket{X}=0$, if \ $\exists N\in\oZ,~
N\leq n :  F_{N}\,F_{N+1}\,\ldots\,F_{n}\,\ket{X}=0$.  Now, an $\N2$ module
$\smU$ is said to belong to the topological $\N2$ category $\TOP$ if, for any
element $\ket{X}$ of $\smU$, $\forall n\in\oZ$
$$\!\!
\BA{ll}
{\rm either}&\!\!\ldots\,Q_{n-3}\,Q_{n-2}\,Q_{n-1}\,Q_{n}\,\ket X=0\\
{\rm or}    &\!\!\ldots\,G_{-n-4}\,G_{-n-3}\,G_{-n-2}\,G_{-n-1}\,\ket X=0
\EA
$$
This condition works by excluding those diagrams that have no `cusps' and
thus, {\it being wider than the diagram~\req{newdiagram2}\/}, necessarily
intersect the edge, at which point the fermionic chain terminates.

Positions of topological singular vectors can be obtained~\cite{[S-sing]} from
the analysis of the Ka\v c determinant \cite{[BFK]}:
A topological singular vector exists in the topological Verma module
$\mV_{h,t}$ iff either $h=\hplus(r,s,t)$ or $h=\hminus(r,s,t)$, where
\BE\kern-4pt\new\BA{ll}
\hplus(r,s,t)=-\frac{r-1}{t}+s-1\\
\hminus(r,s,t)=\frac{r+1}{t}-s
\EA ~r,s\in\oN
\label{n2singcon}\EE

I now introduce two operators $g(\mu,\nu)$ and $q(\mu,\nu)$, with
$\mu,\nu\in\oC$, that represent the action of two ``$\N2$ Weyl group"
generators when $\mu$ and $\nu$ are special (see \cite{[ST3]} for the
details).  These operators act on the plane $t=\const$ as follows
\BE\new\BA{l}
g(ht+\theta-1,\theta-1)\,:\,\kettop{h,t;\theta}\mapsto\\
\quad\kettop{\frac{2}{t}-h,t;ht+\theta-1},\\
q(-(h+1)t-\theta+1,-\theta-1)\,:\,\kettop{h,t;\theta}\mapsto\kern-19pt\\
\quad\kettop{\frac{2}{t}-2-h,t;(h+1)t+\theta-1}\,.
\label{n2weylaction}\EA\EE
The action of $g(a,b)$ and $q(a,b)$ on highest-weight vectors can be extended
to the corresponding topological Verma modules, which allows one to
explicitly construct singular vectors in the topological Verma
modules~$\mV_{\htop^\pm(r,s,t),t}$:
\begin{eqnarray}
&&\kern-20pt\ket{E(r,s,t)}^+= g(-r,(s-1)t-1)\cdot\nonumber\\
&&\kern-22pt\quad q(-(s-1)t,r-1-t)\ldots{}g((s-2)t-r,t-1)\cdot\kern-30pt
\nonumber\\
&&\kern-22pt\qquad q(-t,r-1-t(s-1))\cdot\label{Tplus}\\
&&\kern-22pt\qquad\quad g((s-1)t-r,-1)\,\ket{\htop^+(r,s,t),t}_{\rm top}\,,
\nonumber
\end{eqnarray}
\BE\kern-5pt\new\BA{l}
\ket{E(r,s,t)}^-=q(-r, (s-1) t - 1)\cdot\\
\quad g(-(s-1)t, r - t - 1)\ldots q((s-2) t - r, t-1)\cdot\kern-30pt\\
\qquad g(-t, r - (s-1) t - 1)\cdot\\
\qquad\quad q((s-1) t - r, -1)\,\ket{\htop^-(r,s,t),t}_{\rm top}
\EA\label{Tminus}\EE

One also introduces the `massive' $\N2$ Verma modules
$\smW_{h,\ell,t;\theta}$, in which the \hw{} states satisfy the following
annihilation and eigenvalue conditions:
\BE\new\BA{rcl}
\multicolumn{3}{l}{L_m\ket{h,\ell,t;\theta}=H_m\ket{h,\ell,t;\theta}=0\,,~
m\geq1,}\\
Q_\lambda\ket{h,\ell,t;\theta}&=&0,~\lambda\in-\theta+\oN\\
G_\nu\ket{h,\ell,t;\theta}&=&0,~\nu=\theta+\oN_0\\
(H_0+\frac{\ctop}{3}\theta)\,\ket{h,\ell,t;\theta}&=&
h\,\ket{h,\ell,t;\theta},\\
\multicolumn{3}{l}{(L_0+\theta H_0+\frac{\ctop}{6}(\theta^2+\theta))
\,\ket{h,\ell,t;\theta}=\ell\ket{h,\ell,t;\theta}\,.\kern-29pt{}}
\EA\EE
The {\it untwisted\/} module, freely generated from $\ket{h,\ell,
t}\equiv\ket{h,\ell, t;0}$, is denoted by~$\mU_{h,\ell,t}$.  They are called
massive because of their property to have a dimension $\ell$ generally
different from zero.

Extremal diagrams of massive Verma modules have the form (in the
untwisted case for simplicity)
\BE
\unitlength=.8mm
\begin{picture}(80,40)
\put(-25,-3){
\put(50.00,10.00){
        \put(00.00,00.00){$\bullet$}
        \put(10.00,20.00){$\bullet$}
        \put(10.00,20.00){$\bullet$}
        \put(20.00,30.00){$\bullet$}
        \put(30.00,30.00){$\bullet$}
        \put(40.00,20.00){$\bullet$}
        \put(50.00,00.00){$\bullet$}
        \put(9.80,19.00){\vector(-1,-2){8}}
        \put(19.20,30.00){\vector(-1,-1){7}}
        \put(22.20,31.50){\vector(1,0){7}}
        \put(32.50,30.00){\vector(1,-1){7}}
        \put(42.00,19.00){\vector(1,-2){8}}
        \put(-2.00,13.00){${}_{G_{-2}}$}
        \put(07.00,26.50){${}_{G_{-1}}$}
        \put(23.90,33.50){${}_{Q_{0}}$}
        \put(37.00,27.50){${}_{Q_{-1}}$}
        \put(47.00,13.00){${}_{Q_{-2}}$}
        \put(-23.00,00.00){${}_{\ket{h_{-2},\ell_{-2},t;-2}}$}
        \put(-14.00,21.00){${}_{\ket{h_{-1},\ell_{-1},t;-1}}$}
        \put(11.00,32.00){${}_{\ket{h_,\ell,t}}$}
        \put(33.50,32.00){${}_{\ket{h_{1},\ell_{1},t;1}}$}
        \put(43.00,21.00){${}_{\ket{h_{2},\ell_{2},t;2}}$}
        \put(53.00,00.00){${}_{\ket{h_{3},\ell_{3},t;3}}$}
        \put(00.50,-06.00){$\vdots$}
        \put(50.50,-06.00){$\vdots$}
        }
}
\end{picture}
\label{massdiagramdouble}\EE
One can map back towards $\ket{h,\ell,t}$ as
\BE\kern-4pt\new
\arraycolsep=2pt\BA{rcl}
Q_{-\theta}\ket{h,\ell,t;\theta}&=&
2\ell\ket{h-\frac{2}{t},\ell+h-\frac{2}{t},t;\theta+1},\phantom{_{\int}}
\kern-30pt\\
G_{\theta-1}\,\ket{h,\ell,t;\theta}&=&
2(\ell-h)\ket{h+\frac{2}{t},\ell-h,t;\theta-1}\phantom{^{\int}}\kern-30pt
\EA\label{mapback}\EE
for $\theta<0$ and $\theta>0$ respectively; this may lead to the vanishing
result, which gives the conditions for the so-called `charged' singular
vectors~\cite{[BFK]} to appear:
\begin{thm}
A massive Verma module $\mU_{h,\ell,t}$ contains a twisted topological
Verma submodule iff $\ell=\theel_{\rm ch}(r,h,t)$, where
\BE
\theel_{\rm ch}(r,h,t)=r(h+\frac{r-1}{t})\,,\quad r\in\oZ\setminus\{0\};
\label{3rdconditions}
\EE
The corresponding singular vector reads
$$
\kern-3pt\ket{E(r,h,t)}_{\rm ch}\!=\!\left\{\kern-6pt\new\BA{l}
Q_{r}\ldots Q_0\ket{h,\theel_{\rm ch}(r,h,t),t},~r\leq-1\\
G_{-r}\ldots G_{-1}\ket{h,\theel_{\rm ch}(r,h,t),t},r\geq1
\EA\right.
\label{thirdE}
$$
\end{thm}

These `charged' singular vectors are analogous to Verma points encountered in
extremal diagrams of relaxed $\tSL2$ modules, see~\req{withVerma}. Moreover,
extremal diagrams of $\N2$ modules are nothing but a deformation (of straight
lines into parabolas) of $\tSL2$ extremal diagrams. This follows from the
statement of equivalence of categories given in Sect.~4, or can be derived by
a case-by-case analysis of all possible branchings of extremal diagrams for
each of the two algebras; in view of the advertised result, we omit the
analysis of $\N2$ singular vectors in the cases when extremal diagrams branch
and several singular vectors coexist in the module.

\medskip

The above condition for the fermionic chains to terminate is {\it not\/}
satisfied for massive Verma modules.  Instead, a fermionic chain terminates
whenever it is yet wider than the diagram \req{massdiagramdouble}, i.e.\ it
contains two or more arrows along the same straight line. Thus, an $\N2$
module $\smU$ belongs to the category $\MHW$ of $\N2$ modules if,
for any element $\ket{X}$ of $\smU$, $\forall n\in\oZ$,
$$
\BA{ll}
{\rm either}&\ldots\,Q_{n-3}\,Q_{n-2}\,Q_{n-1}\,Q_{n}\,\ket X=0\\
{\rm or}    &\ldots\,G_{-n-4}\,G_{-n-3}\,G_{-n-2}\,G_{-n}\,\ket X=0.
\EA
$$

\section{Kazama--Suzuki and related mappings}
The simplest KS construction uses
a couple of spin-1 $BC$ ghosts, which allows one to build up the topological
algebra generators as~\cite{[DvPYZ],[KS],[Le],[EHy]}:
\begin{eqnarray}
&&\kern-20pt Q=CJ^+,~G=\frac{2}{k+2}BJ^-,~H=\frac{k}{k+2}BC-
\frac{2}{k+2}J^0,\kern-10pt\nonumber\\
&&\kern-20pt T=\frac{1}{k+2}(J^+J^-)-\frac{k}{k+2}B\d C-\frac{2}{k+2}BCJ^0\,.
\label{QGsl}\label{Tsl}
\end{eqnarray}
Generators \req{Tsl} close to the algebra \req{topalgebra}, the central
elements being related by $\ctop={3k\over k+2}$.  Thus, eqs.~\req{Tsl}
define a mapping
\BE
\FKS:\cA\to\cU\tSL2_k\tensor\,[BC],\label{frttosl(2)}
\EE
where $\cA$ is the $\N2$ algebra \req{topalgebra} and $\cU$ denotes the
universal enveloping (and $[BC]$ is the free fermion theory).

The KS mapping produces also a bosonic current
\BE
I^+=\sqrt{\frac{2}{k+2}}(BC+J^0)\label{Heisenberg}
\EE
whose modes commute with the $\N2$ generators \req{Tsl}.  Its
\emt\ reads
\BE
T^+=\half\,(I^+)^2-\frac{1}{\sqrt{2(k+2)}}\d\,I^+
\EE
Expanding as $I^+(z)=\sum_{n\in\oZ}I^+_nz^{-n-1}$, I get a Heisenberg
algebra $[I^+_m,\,I^+_n]=m\delta_{m+n,0}$.
I define a module $\mH_p^+$ over the Heisenberg algebra by the
highest-weight conditions
\BE
I_{\geq1}\,\ket{p}^+=0,~ I_0\,\ket{p}^+=p\,\ket{p}^+
\label{Heishw}\EE
Under the KS mapping~\req{Tsl}, one has the identities
\BE
T_{\rm Sug}+T_{\rm GH}=T + T^+
\EE
(where $T$ is the \emt~\req{Tsl}), and
\BE
J^0-BC=-2 H+\frac{k-2}{\sqrt{2(k+2)}}\,I^+\,.
\EE

A mapping in the inverse direction to the KS mapping is constructed using the
above `antifermions' $\psi=e^\phi$, $\spsi=e^{-\phi}$:
\BE\new\BA{l}
J^+= Q\psi,~ J^-=\frac{3}{3-\ctop}\, G\spsi,\\
J^0=-\frac{3}{3-\ctop}\,H+\frac{k}{2}\,\d\phi
\EA\label{invKS}\EE
For $\ctop\neq3$, generators \req{invKS} close to the affine $\SL2$ algebra
of the level $k={2\ctop\over3-\ctop}$ where $\ctop$ is the $\N2$ central
charge. Thus, Eqs.~\req{invKS} define a mapping
\BE
\FKS^{-1}:\tSL2\to\cU\,\cA\tensor[\psi\,\psi^*]\,.
\EE

One also has a free scalar with signature $-1$, whose modes commute
with the $\tSL2$ generators:
\BE
I^-=\sqrt{\frac{3}{3-\ctop}}(H-\d\phi)\,.
\label{dF}\EE
The modes of $I^-(z)=\sum^{\infty}_{n=-\infty}I^-_n z^{-n-1}$ generate a
Heisenberg algebra.  The module $\mH^-_q$ is defined as a Verma module over
the Heisenberg algebra with the highest-weight vector defined by
\BE
I^-_n\ket{q}^-=0,\qquad n\geq1,~ I^-_0\ket{q}^-=q\ket{q}^-
\EE
Under the anti-KS mapping \req{invKS} one has the identities
\BE
T+T_\phi=T^{\rm Sug}+T^-
\EE
(where $T$ is the \emt\ of the $\N2$ algebra and $T_\phi$ is the \emt\ of the
Liouville system \req{pseudoemt}), and
\BE
-2H+\d\phi=J^0+\L(\frac{k}{2}-1\R)\sqrt{\frac{k+2}{2}}I^-.
\EE

The composition $F_{\rm KS}^{-1}\circ F_{\rm KS}$
maps the $\tSL2$ algebra into an $\tSL2$ algebra in the tensor product
$\cU\,\tSL2\tensor[BC]\tensor[\psi\,\psi^*]$
\BE\new\BA{l}
\bar J^+=J^+e^{\phi}\,C,~\bar J^-=J^-e^{-\phi}B,\\
\bar J^0=J^0+\frac{k}{2}(\d\phi-BC)\,.
\label{sl2back}\EA\EE

The same happens with the $\N2$ algebra under the action of $F_{\rm KS}\circ
F_{\rm KS}^{-1}$, which maps the $\N2$ algebra $\cA$ into an $\N2$ algebra in
the tensor product $\cU\,\cA\tensor[BC]\tensor[\psi\,\psi^*]$
\BE\new\BA{rcl}
\bar Q&=&Q e^{\phi}C,~\bar G= G e^{-\phi}B,\\
\bar H&=&H+\frac{k}{k+2}(BC-\d\phi),\\
\bar T&=&T+H(BC-\d\phi)+{}\\
\multicolumn{3}{l}{\qquad\frac{k}{2(k+2)}
\L((\d\phi)^2-2\d\phi\,BC+\d^2\phi - 2 B\,\d C\R)\,.\kern-24pt}
\EA\label{N2back}\EE

The above identities allow one to deduce
\begin{thm}\label{mainthm}\mbox{}\nopagebreak

{\rm I.} The KS mapping induces an isomorphism of $\N2$ representations
$$
\smM_{j,k;\theta}\tensor\Omega\,{}\approx \bigoplus\limits_{\lambda\in\oZ}\,
\smV_{{-2j\over k+2},k+2;\lambda-\theta}\tensor
\mH^+_{\sqrt{{2\over k+2}}(j-\frac{k}{2}\theta-\lambda)}
$$
where on the LHS the $\N2$ algebra acts by the generators
\req{QGsl}, while on the RHS it acts on $\smV_{{-2j\over
k+2},k+2;\lambda-\theta}$ as on its (twisted topological) Verma module.

{\rm II.} The anti-KS mapping induces an isomorphism of $\tSL2$ representations
$$
\smV_{h,t;\theta}\tensor\Xi\approx\bigoplus\limits_{n\in\oZ}\,
\smM_{-\frac{t}{2}h,t-2;n-\theta}\tensor
\mH^-_{\sqrt{\frac{t}{2}}(h-\frac{t-2}{t}\theta+n)}
$$
where on the LHS the $\tSL2$ algebra acts by the generators \req{invKS}, while
on the RHS it acts on $\smM_{-\frac{t}{2}h,t-2;n-\theta}$ as on its twisted
Verma module.
\end{thm}

As a corollary, observe that singular vectors in $\tSL2$ Verma modules and in
topological $\N2$ Verma modules occur (or do not occur) simultaneously.

A `relaxed' version of the above result reads
\begin{thm}\label{relaxedmainthm}\mbox{}

{\rm I.} The KS mapping induces an isomorphism of $\N2$ representations
\BE\BA{l}
\smR_{j,\Lambda,k;\theta}\tensor\Omega\,{}\approx \bigoplus_{\lambda\in\oZ}\,
\smW_{{-2j\over k+2},\frac{\Lambda}{k+2},k+2;\lambda-\theta}\tensor\kern-10pt\\
\qquad\qquad\mH^+_{\sqrt{{2\over k+2}}(j-\frac{k}{2}\theta-\lambda)}
\EA\label{relaxedidspaces}
\EE
where on the LHS the $\N2$ algebra acts by the
generators~\req{Tsl}, while on the RHS it acts naturally on
$\smW_{{-2j\over k+2},\ell,k+2;\lambda-\theta}$ as on a twisted massive
Verma module.

{\rm II.} The anti-KS mapping induces an isomorphism of $\tSL2$ representations
\BE\BA{l}
\smW_{h,\ell,t;\theta}\tensor\Xi\approx\bigoplus\limits_{n\in\oZ}\,
\smR_{-\frac{t}{2}h,\Lambda(\ell),t-2;n-\theta}\tensor{}\kern-20pt\\
\qquad\qquad\mH^-_{\sqrt{\frac{t}{2}}(h-\frac{t-2}{t}\theta+n)}
\EA\label{relaxedidspaces2}
\EE
where on the LHS the $\tSL2$ algebra acts by generators \req{invKS}, while on
the RHS it acts naturally on $\smR_{-\frac{t}{2}h,t-2;n-\theta}$ as on a
twisted relaxed Verma module.
\end{thm}
This is illustrated by the following diagram:
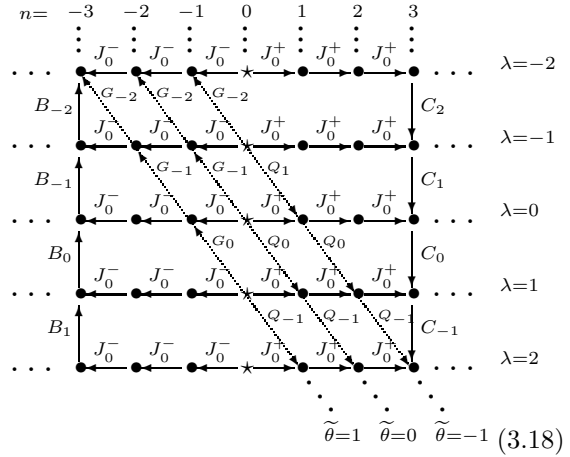
\BE
\unitlength=.7pt
\begin{picture}(250,240)
\put(40,0){%
        \put(0,120){
        \put(-30,107){${}^{n={}}$}
        \put(-10,95){
        \put(6,13){${}^{-3}$}
        \put(0,0){
          \put(9.5,-5){\Large $\cdot$}
          \put(9.5,1){\Large $\cdot$}
          \put(9.5,7){\Large $\cdot$}
        }
        }
        \put(20,95){
        \put(6,13){${}^{-2}$}
        \put(0,0){
          \put(9.5,-5){\Large $\cdot$}
          \put(9.5,1){\Large $\cdot$}
          \put(9.5,7){\Large $\cdot$}
        }
        }
        \put(50,95){
        \put(6,13){${}^{-1}$}
        \put(0,0){
          \put(9.5,-5){\Large $\cdot$}
          \put(9.5,1){\Large $\cdot$}
          \put(9.5,7){\Large $\cdot$}
        }
        }
        \put(80,95){
        \put(10.3,13){${}^{0}$}
        \put(0,0){
          \put(9.5,-5){\Large $\cdot$}
          \put(9.5,1){\Large $\cdot$}
          \put(9.5,7){\Large $\cdot$}
        }
        }
        \put(110,95){
        \put(10.3,13){${}^{1}$}
        \put(0,0){
          \put(9.5,-5){\Large $\cdot$}
          \put(9.5,1){\Large $\cdot$}
          \put(9.5,7){\Large $\cdot$}
        }
        }
        \put(140,95){
        \put(10.3,13){${}^{2}$}
        \put(0,0){
          \put(9.5,-5){\Large $\cdot$}
          \put(9.5,1){\Large $\cdot$}
          \put(9.5,7){\Large $\cdot$}
        }
        }
        \put(170,95){
        \put(10.3,13){${}^{3}$}
        \put(0,0){
          \put(9.5,-5){\Large $\cdot$}
          \put(9.5,1){\Large $\cdot$}
          \put(9.5,7){\Large $\cdot$}
        }
        }
        %
        \put(2.4,7){\vector(0,1){30}}
        \put(187,17){${}^{C_{1}}$}
        \put(2.4,47){\vector(0,1){30}}
        \put(187,57){${}^{C_{2}}$}
        \put(2.4,-33){\vector(0,1){30}}
        \put(187,-23){${}^{C_{0}}$}
        \put(2.4,-73){\vector(0,1){30}}
        \put(187,-63){${}^{C_{-1}}$}
        \put(182.5,37){\vector(0,-1){30}}
        \put(-22,17){${}^{B_{-1}}$}
        \put(182.5,77){\vector(0,-1){30}}
        \put(-22,57){${}^{B_{-2}}$}
        \put(182.5,-3){\vector(0,-1){30}}
        \put(-15,-23){${}^{B_{0}}$}
        \put(182.5,-43){\vector(0,-1){30}}
        \put(-15,-63){${}^{B_{1}}$}
        %
        \put(60,0){
        \put(-30,0){
          \put(154,-90){\Large $\cdot$}
          \put(160,-97){\Large $\cdot$}
          \put(166,-104){\Large $\cdot$}
          \put(165,-110){${}_{\widetilde\theta=-1}$}
        }
        \bezier{100}(5,80)(62.5,2)(120,-74)
        \put(5,80){\vector(-2,3){1}}
        \put(59.5,7){\vector(2,-3){1}}
        \put(90.7,-34.7){\vector(2,-3){1}}
        \put(120,-74){\vector(2,-3){1}}
        \put(44,30){${}_{{}^{Q_{1}}}$}
        \put(14,70){${}_{{}^{G_{-2}}}$}
        \put(74,-10){${}_{{}^{Q_{0}}}$}
        \put(104,-50){${}_{{}^{Q_{-1}}}$}
        }
        \put(30,0){
        \put(-30,0){
          \put(154,-90){\Large $\cdot$}
          \put(160,-97){\Large $\cdot$}
          \put(166,-104){\Large $\cdot$}
          \put(165,-110){${}_{\widetilde\theta=0}$}
        }
        \bezier{100}(5,80)(62.5,2)(120,-74)
        \put(5,80){\vector(-2,3){1}}
        \put(35.5,39){\vector(-2,3){1}}
        \put(90.7,-34.7){\vector(2,-3){1}}
        \put(120,-74){\vector(2,-3){1}}
        \put(44,30){${}_{{}^{G_{-1}}}$}
        \put(14,70){${}_{{}^{G_{-2}}}$}
        \put(74,-10){${}_{{}^{Q_{0}}}$}
        \put(104,-50){${}_{{}^{Q_{-1}}}$}
        }
        \put(0,0){
        \put(-30,0){
          \put(154,-90){\Large $\cdot$}
          \put(160,-97){\Large $\cdot$}
          \put(166,-104){\Large $\cdot$}
          \put(165,-110){${}_{\widetilde\theta=1}$}
        }
        \bezier{100}(5,80)(62.5,2)(120,-74)
        \put(5,80){\vector(-2,3){1}}
        \put(35.5,39){\vector(-2,3){1}}
        \put(65.7,-1.3){\vector(-2,3){1}}
        \put(120,-74){\vector(2,-3){1}}
        \put(44,30){${}_{{}^{G_{-1}}}$}
        \put(14,70){${}_{{}^{G_{-2}}}$}
        \put(74,-10){${}_{{}^{G_{0}}}$}
        \put(104,-50){${}_{{}^{Q_{-1}}}$}
        }
        %
        %
        \put(230,0){${}^{\lambda=0}$}
        \put(0,0){
        \put(-35,2){\Large $\ldots$}
        \put(0,0){$\bullet$}
        \put(10,5){${}^{J^-_0}$}
        \put(28,3){\vector(-1,0){22}}
        \put(30,0){$\bullet$}
        \put(40,5){${}^{J^-_0}$}
        \put(58,3){\vector(-1,0){22}}
        \put(60,0){$\bullet$}
        \put(70,5){${}^{J^-_0}$}
        \put(88,3){\vector(-1,0){22}}
        \put(90,0){$\star$}
        \put(100,5){${}^{J^+_0}$}
        \put(97,3){\vector(1,0){22}}
        \put(120,0){$\bullet$}
        \put(130,5){${}^{J^+_0}$}
        \put(127,3){\vector(1,0){22}}
        \put(150,0){$\bullet$}
        \put(160,5){${}^{J^+_0}$}
        \put(157,3){\vector(1,0){22}}
        \put(180,0){$\bullet$}
        \put(193,2){\Large $\ldots$}
        }
        %
        %
        \put(230,40){${}^{\lambda=-1}$}
        \put(0,40){
        \put(-35,2){\Large $\ldots$}
        \put(0,0){$\bullet$}
        \put(10,5){${}^{J^-_0}$}
        \put(28,3){\vector(-1,0){22}}
        \put(30,0){$\bullet$}
        \put(40,5){${}^{J^-_0}$}
        \put(58,3){\vector(-1,0){22}}
        \put(60,0){$\bullet$}
        \put(70,5){${}^{J^-_0}$}
        \put(88,3){\vector(-1,0){22}}
        \put(90,0){$\star$}
        \put(100,5){${}^{J^+_0}$}
        \put(97,3){\vector(1,0){22}}
        \put(120,0){$\bullet$}
        \put(130,5){${}^{J^+_0}$}
        \put(127,3){\vector(1,0){22}}
        \put(150,0){$\bullet$}
        \put(160,5){${}^{J^+_0}$}
        \put(157,3){\vector(1,0){22}}
        \put(180,0){$\bullet$}
        \put(193,2){\Large $\ldots$}
        }
        %
        \put(230,-40){${}^{\lambda=1}$}
        \put(0,-40){
        \put(-35,2){\Large $\ldots$}
        \put(0,0){$\bullet$}
        \put(10,5){${}^{J^-_0}$}
        \put(28,3){\vector(-1,0){22}}
        \put(30,0){$\bullet$}
        \put(40,5){${}^{J^-_0}$}
        \put(58,3){\vector(-1,0){22}}
        \put(60,0){$\bullet$}
        \put(70,5){${}^{J^-_0}$}
        \put(88,3){\vector(-1,0){22}}
        \put(90,0){$\star$}
        \put(100,5){${}^{J^+_0}$}
        \put(97,3){\vector(1,0){22}}
        \put(120,0){$\bullet$}
        \put(130,5){${}^{J^+_0}$}
        \put(127,3){\vector(1,0){22}}
        \put(150,0){$\bullet$}
        \put(160,5){${}^{J^+_0}$}
        \put(157,3){\vector(1,0){22}}
        \put(180,0){$\bullet$}
        \put(193,2){\Large $\ldots$}
        }
        %
        %
        \put(230,80){${}^{\lambda=-2}$}
        \put(0,80){
        \put(-35,2){\Large $\ldots$}
        \put(0,0){$\bullet$}
        \put(10,5){${}^{J^-_0}$}
        \put(28,3){\vector(-1,0){22}}
        \put(30,0){$\bullet$}
        \put(40,5){${}^{J^-_0}$}
        \put(58,3){\vector(-1,0){22}}
        \put(60,0){$\bullet$}
        \put(70,5){${}^{J^-_0}$}
        \put(88,3){\vector(-1,0){22}}
        \put(90,0){$\star$}
        \put(100,5){${}^{J^+_0}$}
        \put(97,3){\vector(1,0){22}}
        \put(120,0){$\bullet$}
        \put(130,5){${}^{J^+_0}$}
        \put(127,3){\vector(1,0){22}}
        \put(150,0){$\bullet$}
        \put(160,5){${}^{J^+_0}$}
        \put(157,3){\vector(1,0){22}}
        \put(180,0){$\bullet$}
        \put(193,2){\Large $\ldots$}
        }
        %
        %
        \put(230,-80){${}^{\lambda=2}$}
        \put(0,-80){
        \put(-35,2){\Large $\ldots$}
        \put(0,0){$\bullet$}
        \put(10,5){${}^{J^-_0}$}
        \put(28,3){\vector(-1,0){22}}
        \put(30,0){$\bullet$}
        \put(40,5){${}^{J^-_0}$}
        \put(58,3){\vector(-1,0){22}}
        \put(60,0){$\bullet$}
        \put(70,5){${}^{J^-_0}$}
        \put(88,3){\vector(-1,0){22}}
        \put(90,0){$\star$}
        \put(100,5){${}^{J^+_0}$}
        \put(97,3){\vector(1,0){22}}
        \put(120,0){$\bullet$}
        \put(130,5){${}^{J^+_0}$}
        \put(127,3){\vector(1,0){22}}
        \put(150,0){$\bullet$}
        \put(160,5){${}^{J^+_0}$}
        \put(157,3){\vector(1,0){22}}
        \put(180,0){$\bullet$}
        \put(193,2){\Large $\ldots$}
        }
        }
}
\end{picture}
\label{lattice}
\EE
This describes the case of $\theta=0$ in \req{relaxedidspaces} (which is the
$\tSL2$ `spectral' parameter, not to be confused with the one labelling
twisted $\N2$ modules that are also present in the diagram; the $\N2$
spectral parameter is denoted by $\widetilde\theta$ here). The diagram
represents the tensor product of the relaxed Verma module extremal diagram
\req{floor} with a ghost extremal diagram. The latter consists of the ghost
vacua~\req{lambdavac} in different pictures.  As we have chosen $\theta=0$,
the $\tSL2$ arrows are horizontal, as in~\req{floor}, while the ghost ones
are shown as vertical.  For simplicity, the ghost ($B$ and $C$) arrows are
shown explicitly only in two columns.  Since the different pictures
\req{lambdavac} in the free-fermion system are all equivalent, all the
vertical arrows are invertible, however we have separated the $B$ and $C$
arrows in the diagram, trying to keep it readable. Further, the~$\bullet$s
denote~$\ketSL{j,\Lambda,k|n}\tensor\ketGH{\lambda}$, the values of $n$ being
written in the upper row and those of $\lambda$, in the right column.
The~$\star$s denote~$\ketSL{j,\Lambda,k}\tensor\ketGH{\lambda}$. {\it The
dotted lines are precisely the extremal diagrams~\req{massdiagramdouble} of
massive Verma modules~$\smW_{h,\ell,t;\widetilde\theta}$ from the RHS
of~\req{relaxedidspaces}, viewed from above\/}.

\section{Categorial equivalences}
The above theorems suggest that one extends the mappings $F_{\rm KS}$ and
$F^{-1}_{\rm KS}$ to a functor that would establish the correspondence
between {\it categories\/} of $\tSL2$ and $\N2$ modules.  However, a
difficulty in defining such a functor can be seen already when attempting to
relate modules $\smM_{j,k;\theta}$ and $\smV_{h,t;\theta'}$ for fixed
$\theta$ and $\theta'$. While on the $\tSL2$ side any submodule of
$\smM_{j,k;\theta}$ is again a twisted Verma module with the same value
of~$\theta$, this is not so on the $\N2$ side, where submodules of a
(twisted) topological Verma module are the twisted topological Verma modules
with {\it different\/} values of the `spectral' parameter~$\theta$.

Thus, an equivalence between some categories of $\tSL2$ and $\N2$ modules can
only be established for those categories that effectively allow for a
factorization with respect to the spectral flow.  These categories can be
defined as follows.  Consider the objects that are infinite chains
$\left(\smM_{j,k;\theta}\right)_{\theta\in\oZ}$, where $\smM_{j,k;\theta}$
are twisted $\tSL2$ Verma modules.  As a morphisms between
$\left(\smM_{j,k;\theta}\right)_{\theta\in\oZ}$ and
$\left({\smM}_{j',k';\theta}\right)_{\theta\in\oZ}$, take any Verma module
morphism $\smM_{j,k;\theta_1}\to {\smM}_{j',k';\theta_2}$.  Call this
category the $\tSL2$ Verma chain category $\CVER$.  The meaning of the
definition of morphisms of chains is that, given a morphism between any two
modules, one spreads it over the entire chains by spectral flow transforms.

On the $N=2$ side, the topological Verma chain category $\CTVER$ is defined
similarly: one takes chains of twisted topological Verma modules, every such
chain consisting of twisted topological Verma modules with all
$\theta\in\oZ$.  Morphisms of the chains are defined similarly to the $\tSL2$
case\footnote{it is important to note that any submodule of a (twisted)
topological Verma module is a twisted topological Verma module.}.

To define a functor relating such chains, I first construct correspondences
between individual modules in the chains.  Given a topological Verma module
$\smV_{h,t;\theta}$, and an arbitrary $\theta'\in\oZ$, take the Heisenberg
modules $\mH^+_{-\sqrt{{2\over t}}(\frac{t}{2}j+\frac{t}{2}\theta'-\theta)}$
and construct
\BE\BA{l}
\smV_{h,t;\theta}\tensor
\mH^+_{-\sqrt{{2\over t}}(\frac{t}{2}j+\frac{t}{2}\theta'-\theta)}
\oplus{}\\~{}\bigoplus\limits_{m\in\oZ,m\neq 0}\,
\smV_{h,t;\theta+m}\tensor
\mH^+_{-\sqrt{{2\over t}}(\frac{t}{2}j+\frac{t}{2}\theta'-\theta+m)}\kern-30pt
\EA
\label{FKS}
\EE
This is isomorphic to the tensor product of an
$\tSL2$ Verma module $\smM_{-\frac{t}{2}h,t-2;\theta'}$ with a ghost module.
Define the result of applying $F_{\rm KS}(\theta,\theta')$
to $\smV_{h,t;\theta}$ to be the module $\smM_{-\frac{t}{2}h,t-2;\theta'}$:
$$
F_{\rm KS}(\theta,\theta')\,:\,\smV_{h,t;\theta}
\leadsto
\smM_{-\frac{t}{2}h,t-2;\theta'},~\theta,\theta'\in\oZ\,.
$$
Similarly,
$$
F^{-1}_{\rm KS}(\theta,\theta')\,:\,\smM_{j,k;\theta}
\leadsto
\smV_{-\frac{2}{k+2}j,k+2;\theta'},~\theta,\theta'\in\oZ
$$
is defined as follows. Given a Verma module $\smM_{j,k;\theta}$ and
$\theta'\in\oZ$, construct the sum
\BE\BA{l}
\smM_{j,k;\theta}\tensor
\mH^-_{-\sqrt{\frac{2}{k+2}}(j+\theta'-\frac{k+2}{2}\theta)}
\oplus\\~{}\bigoplus\limits_{n\in\oZ,n\neq0}\,
\smM_{j+\frac{k}{2}n,k;\theta+n}\tensor
\mH^-_{-\sqrt{\frac{2}{k+2}}(j+\theta'-\frac{k+2}{2}\theta+n)}\kern-30pt
\EA
\label{IFKS}
\EE
which is isomorphic to the module $\smV_{-\frac{2}{k+2}j,k+2;\theta'}$
tensored with a module of antifermions.  This twisted topological $\N2$
module is by definition the result of applying $\FKS^{-1}(\theta,\theta')$ to
$\smM_{j,k;\theta}$.

While $F_{\rm KS}$ and $F_{\rm KS}^{-1}$ depend on chosen $\theta$ and
$\theta'$, the $\theta$-dependence disappears when applied to the elements
of $\CVER$ and $\CTVER$:
$$\new\BA{rclcl}
\FKS&:&\CTVER&\leadsto&\CVER\\
\FKS^{-1}&:&\CVER&\leadsto&\CTVER
\EA
$$
Evidently, the composition of $F_{\rm KS}$ and $F^{-1}_{\rm KS}$ maps each
chain of twisted Verma modules into an isomorphic chain. Therefore, $\FKS$
and $\FKS^{-1}$ would become the direct and inverse functors as soon as I
define how $F_{\rm KS}$ and $F^{-1}_{\rm KS}$ act on morphisms.

Recall that in Verma module categories, morphisms are naturally identified
with singular vectors.  As can be seen from the above isomorphisms, an
$\tSL2$ singular vector exists in a twisted Verma module $\smM_{j,k;\theta}$
(consequently in all those with $\theta\mapsto\theta+n$, $n\in\oZ$) iff a
topological singular vector exists in at least one
twisted topological Verma module $\smV_{{-2j\over k+2},k+2;m}$, $m\in\oZ$.
An explicit mapping between the `building blocks'
of the corresponding singular vectors is described as follows:
The KS mapping induces a correspondence between the `continued' objects,
$(J^-_{-\theta})^{\nu-\mu+1}$ and $(J^+_{\theta-1})^{\nu-\mu+1}$ on the one
hand, and $g(\mu,\nu)$, $q(\mu,\nu)$ on the other hand, which act on the
respective highest weights as shown in \req{sl2weylaction}
and \req{n2weylaction}.  The correspondence reads
$$\new\BA{l}
F_{\rm KS}(\theta',\theta)\,:\,g(\mu,\nu)\mapsto
(J^-_{-\theta})^{\nu-\mu+1}b(\mu,\nu),\\
F_{\rm KS}(\theta',\theta)\,:\,q(\mu,\nu)\mapsto
(J^+_{\theta-1})^{\nu-\mu+1}c(\mu,\nu),\\
F^{-1}_{\rm KS}(\theta,\theta')\,:\,(J^-_{-\theta})^{\nu-\mu+1}\mapsto
g(\mu,\nu)e^{-(\nu-\mu+1)\phi},\\
F^{-1}_{\rm KS}(\theta,\theta')\,:\,(J^+_{\theta-1})^{\nu-\mu+1}\mapsto
q(\mu,\nu)e^{(\nu-\mu+1)\phi}
\EA
$$

This leads to
\begin{thm}
The KS and anti-KS mappings give rise to isomorphisms, denoted again
by $F_{\rm KS}(\theta,\theta')$ and $F^{-1}_{\rm KS}(\theta,\theta')$
respectively, between the (twisted) $\tSL2$ singular vectors \req{mffplus} in
the Verma modules $\smM_{j,k;\theta}$ and the (twists of) $\N2$ singular
vectors \req{Tplus}, \req{Tminus} in the (twisted) topological Verma modules
$\smV_{-\frac{2}{k+2}j,k+2;\theta}$:
\BE\new\BA{l}
F_{\rm KS}(\theta,\theta')\,:\,\ket{E(r,s,k+2)}^{\pm,\theta}
\mapsto{}\\
\kern70pt
\ket{S_\pm^{\rm MFF}(r,s,k)}^{\theta'},
\EA\EE
\BE\new\BA{l}
F^{-1}_{\rm KS}(\theta,\theta')\,:\,
\ket{S_\pm^{\rm MFF}(r,s,k)}^{\theta}
\mapsto\\
\kern70pt\ket{E(r,s,k+2)}^{\pm,\theta'},
\EA\label{KSonvectors}\EE
where $\ket{E(r,s,t)}^{\pm,\theta}$ and $\ket{S_\pm^{\rm
MFF}(r,s,k)}^{\theta'}$ denote the respective singular vectors
transformed by the corresponding spectral flows.
\end{thm}

Evidently, $\FKS(\,\cdot,\,\cdot\,)$ and $\FKS^{-1}(\,\cdot,\,\cdot\,)$
applied to chains of the respective Verma modules take morphisms (between
chains) into morphisms. Thus, finally,
\begin{thm}
The functors $F_{\rm KS}$ and $F^{-1}_{\rm KS}$ are covariant functors which
are inverse to each other and which therefore establish the equivalence of
the Verma chain category on the $\tSL2$ side and the topological Verma chain
category on the $\N2$ side.
\end{thm}

An extension of the above theorems to the relaxed and massive Verma modules
is as follows. For uniformity, I call the Verma \hw{} states in
extremal diagrams of relaxed $\tSL2$ modules (see~\req{withVerma}) the
{\it charged\/} $\tSL2$ singular vector.
\begin{thm}
The KS and anti-KS mappings gives rise to identifications, which we denote
again by $F_{\rm KS}(\theta,\theta')$ and $F^{-1}_{\rm KS}(\theta,\theta')$
respectively, between\\
{\rm 1)}~singular vectors $\Sigma^-(r,s,j,k)$
and $\Sigma^+(r,s,j,k)$ in the relaxed Verma module $\smR_{j,\Lambda,k;\theta}$
and $\N2$ singular vectors in the
twisted massive Verma modules
$\smW_{-\frac{2}{k+2}j,\frac{\Lambda}{k+2},k+2;\theta}$,
and\\
{\rm 2)}~charged $\tSL2$ singular vectors and charged $\N2$
singular vectors:
\BE\new\BA{l}
F_{\rm KS}(\theta,\theta')\,:\,\ket{C(r,j,k)}^\theta
\mapsto
\ket{E(r,h,t)}_{\rm ch}^{\theta'},\\
F^{-1}_{\rm KS}(\theta,\theta')\,:\,
\ket{E(r,h,t)}_{\rm ch}^{\theta}
\mapsto
\ket{C(r,j,k)}^{\theta'},
\EA\EE
where $\ket{\Sigma^\pm(r,s,j,k)}^\theta$ and $\ket{S(r,s,h,t)}^{\mp,\theta'}$
denote the respective singular vectors transformed by the corresponding
spectral flows.
\end{thm}
Thus,
\begin{thm}
The functors $F_{\rm KS}$ and $F^{-1}_{\rm KS}$ are covariant functors which
are inverse to each other and which therefore establish the equivalence of
the relaxed Verma chain category~$\CRVER$
on the $\tSL2$ side and the massive Verma chain
category~$\CMVER$
on the $\N2$ side.
\end{thm}

\pagebreak[3]

To conclude, one of the central notions relating the $\tSL2$ and $\N2$ parts
of the story are the extremal diagrams. As I have already mentioned,
extremal diagrams of $\N2$ modules are nothing but a deformation of $\tSL2$
extremal diagrams when straight lines become `parabolas'.  Yet in the $\tSL2$
case, where extremal diagrams consist of straight lines, it appears more
`obvious' that different states on the same floor are equivalent (generically,
unless charged singular vectors -- {\it Verma\/} \hw{} states -- appear).
In the $\N2$ case, on the other hand, it may be tempting to assign a more
fundamental status to vectors at the top level of (untwisted) extremal
diagrams.  However, working with top-level representatives of $\N2$ extremal
diagrams conceals the true nature of massive $\N2$ modules and the
`topological' structure of some of their submodules.  Back in the $\tSL2$
terms, the top-level representatives of a singular vector would correspond to
those states where the $\Lambda$ parameter reaches an extremum among the
states in the same extremal subdiagram, which does not seem to be of much
practical use.

Thus, there exist two different languages, the $\tSL2$ and $\N2$ ones, to
describe essentially the same structure; a very interesting task is to extend
the dictionary relating representation-theoretic terms for the two algebras
to include a fusion-related vocabulary\footnote{Factoring with respect to the
spectral flows makes the fusion rules isomorphic, thus the difference between
$\tSL2$ and $\N2$ fusion rules is related to different periodicity of the
respective modules under the spectral flows.}.

It is a pleasure to thank the Organizers for a very interesting Conference.
The research described in this work was supported in part by the RFFI Grant
96-01-00725 and by Deutsche Forschungsgemeinschaft under contract 436 RUS
113-29.

\end{document}